\documentclass[galaxies,article,accept,pdftex,moreauthors]{Definitions/mdpi}
\firstpage{1} 
\pubvolume{1}
\issuenum{1}
\articlenumber{0}
\pubyear{2026}
\copyrightyear{2026}
\externaleditor{ Anne M.
Hofmeister 
and Robert E. Criss} 
\datereceived{22 November 2025} 
\daterevised{10 February 2026 } 
\dateaccepted{13 February 2026 } 
\datepublished{25 February 2026 } 
\usepackage{longtable}
      
\Title{SINGLE PARAMETER MODEL FOR GALAXY ROTATION CURVES}
\Author{Sophia N. Cisneros $^{1,\ddagger}$*\orcidA{}, Rich Ott $^{2,\ddagger}$,  Meagan Crowley $^{3,\ddagger}$, Amy Roberts $^{4,\ddagger}$, and Marcus Paz $^{5,\ddagger}$}
\address{$^{1}$ \quad Department of Astronomy, University of Washington, 1410 NE Campus Pkwy, Seattle, 98195, USA;  \url{sofcis94@uw.edu}\\
$^{2}$ \quad Department of Physics, Massachusetts Institute of Technology, 77 Massachusetts Ave, Cambridge,  02139, USA;  \url{rott@alum.mit.edu} \\
$^{3}$ \quad Department of Chemistry, Colorado School of Mines, 1500 Illinois St, Golden, CO 80401, USA;  \url{Mcrowley1@mines.edu}\\
$^{4}$ \quad Department of Physics, University of Colorado Denver, 1201 Larimer St, Denver,  80204, USA; \url{AMY.ROBERTS@ucdenver.edu} \\
$^{5}$ \quad Department of Physics, University of Denver, 2199 S University Blvd, Denver, 80210, USA; \url{marcus.paz@du.edu}}
\corres{Correspondence: \url{sofcis94@uw.edu}}
\secondnote{These authors contributed equally to this work.}
\abstract{One key piece of evidence for dark matter is the rotation-curve problem: the disagreement between measured galactic rotation curves and their luminous mass. 
A novel solution to this problem is presented here, in a model that predicts observed Doppler-shifted spectra based only on the luminous matter estimates and one free model parameter. 
This model is applied to fit the rotation curves of the SPARC sample of 175 galaxies, yielding   mass-to-light ratios, goodness of fit measurements, and the free parameter.
The measured average reduced chi-squared of $2.24$ compares favorably with the Navarro-Frenk-White dark matter model’s average of  $ 4.19$ for the same data, and more galaxies are successfully fit by this model.  
  The model provides a useful formulation linking luminous matter to the observed rotation curves, with the dark matter contribution to galaxies encoded in two transformation terms of the luminous mass.  It 
also offers a lower-parameter characterization of the rotation curve problem,  and    a power law relationship between the model's free parameter and galactic photometric quantities is observed, potentially removing the need for the free parameter.} 

\begin{document}

\section{Introduction}  \label{intro}


At the level of the data, the rotation-curve problem in spiral galaxies arises from the discrepancy between velocities inferred from Doppler-shifted spectra and those derived from luminous mass observations \citep{1970ApJ...159..379R,1978Rubin,Bosma,1985ApJAlbada}. Because both velocities serve as proxies for mass, this discrepancy has been a cornerstone for positing a missing mass component — dark matter. Despite extensive searches, both direct and indirect detection experiments have thus far yielded null results \cite{Cebrian:2022brv,Misiaszek_2024}. Following the Large Hadron Collider's null results for a dark matter candidate and as direct-detection experiments approach the ever-present neutrino background   \citep{neutrino_fog_SNOWMASS}, it is imperative to systematically explore alternative explanations for rotation curve phenomenology.

 Historically, \citet{1973ApJ...186..467O} and \citet{Rub} treated the dark matter problem in spiral galaxies as a distinct physics problem from the cosmological context. In this work, we follow that convention, focusing exclusively on the rotation-curve problem. This choice is motivated by compelling correlations between luminous and dark matter in this context \cite{Bosma,1985ApJAlbada,PSS,SofRub}, which lack a clear physical basis \cite{McGaugh_2000,1978Rubin,PSS,1999McGaugh,2004ApJ...609..652M,1977A&A....54..661T,Ziegler_2002}.
The Universal Rotation Curve (URC), discovered by \citet{salucci} and \citet{SalucciApJ}, shows that when 1,100 galaxy rotation curves are normalized by their scale lengths and plotted together, they naturally fall into three groups: flat rotation curves at the median, large galaxy rotation curves inflecting downwards, and small galaxy rotation curves inflecting upwards. Notably, the Milky Way lies at the median of this distribution. Since there is no a priori reason for the Milky Way to occupy this position, we interpret this as hinting at frame-dependent effects associated with the Milky Way.

 The dark matter paradigm interprets the URC   phenomenology that   galaxies larger than the Milky Way   have    minimal dark matter halos, while  galaxies smaller   than the Milky Way  are dark matter dominated. 
 This interpretation is less than satisfactory, because it contradicts classical gravity on which dark matter theories are built.   Classical gravity requires   that  mass accretion rates are directly proportional to the  initial mass function \cite{10.1093/mnras/stt2403}.  
Accordingly, we take this as motivation to characterize dark matter phenomenology in this context through frame effects associated with the Milky Way, while remaining agnostic about its underlying physics. This approach yields a simpler rotation curve formula with only one free parameter, compared to at least two in NFW and Einasto dark matter models \cite{Iocco_2015}, and successfully fits a greater number of galaxies. Case studies demonstrating the flexibility of this model across all three classes of rotation curves are provided in Appendix~\ref{results:MtoL}.  
 
The necessity  of reducing the parameter space   is directly connected to the study of luminous mass modeling in spiral galaxies.
Luminous mass modeling  is a tricky business,  precisely because  the Doppler-shifted spectra observations — expected to constrain such  models — instead introduced the dark matter problem and at least two more free parameters \cite{Li_2020}. 
Luminous mass models, which underpin rotation curve formulae, already include three parameters to scale the luminous mass components of a spiral galaxy: stellar disk, stellar bulge, and gas halo.  
Embedded in these  three scaling parameters there is a cascade of model-dependent parameter choices \cite{1971ApJS...22..445S}.  
The process of   stellar population synthesis modeling,   which frames  a  galaxy's mass budget, relies on matching the photometric measurement of a galaxy's spectral energy density (SED) to a suite of under-constrained   parameters \cite{Conroy}. 
To extract a meaningful model for a given SED, assumptions must be made regarding the history of star formation and evolution, stellar metallicities and abundances, the states and quantities of gas and dust,   dust attenuation and extinction — and this is by no means an exhaustive list
\cite{Munshi_2013}. 
 Currently no single model can encompass the full range of the  parameter space for galaxies, so models are stitched together, with poorly constrained physics assumptions
  \cite{2013ARA&A..51..393C,doi:10.1139/p10-104,2012ApJ...751...67L}. 
  The SDSS IV MaSTAR Stellar Spectral Library \cite{2019ApJ...883..175Y} has significantly increased coverage and standardization in the field,
and yet   the theoretical modeling of galaxy SEDs remains a convolution of all these effects - making disentanglement difficult and resulting in large uncertainties in the resulting luminous mass model \cite{2016Lelli}.

 The approach presented here offers a promising avenue to refine modeling techniques and constrain galaxy mass estimates. This model does not alter classical physics, but encodes gravitational effects in galaxy frames using standard relativistic tools applied in a novel context.
  We note that   previous studies  which  have ruled out  general relativity in this  context   \citep{Ciotti_2022} did not compare relative galaxy frames as \emph{ratios} of   galaxy gravitational potentials,    one-to-one in radii, as we do. 
  We use the resulting model to    fit the   175 well-studied  spiral galaxy rotation curves in the SPARC sample \cite{2016Lelli} and show how   this   model's   free parameter appears to be highly correlated to a ratio of photometric quantities.

 Other models, including Newtonian mass approaches and geometric analyses of galaxies \cite{doi:10.1139/cjp-2016-0625,galaxies8020036,galaxies8010009},   do not appear as tightly constrained as the present approach. 
 Modified Newtonian Dynamics (MOND) \citet{Milgrom} remains a notable alternative, effectively reproducing rotation curves, but represents a modification of classical gravitational physics rather than a re-expression of frame-dependent effects.
In Sec.~\ref{otros} we describe some of the other models which have been employed in  the spiral galaxy rotation problem, in comparison to our model.

The paper is organized as follows: {\bf Section \ref{sec:dos}} describes the transition from the standard dark matter rotation curve formula to this paper’s form; {\bf Section \ref{sec:methods}} details the methods used to fit spiral galaxy rotation curve data using our model; {\bf Section \ref{sec:analysis}} summarizes the key results; and {\bf Section \ref{sec:conclu}} presents conclusions and future perspectives. Detailed galaxy data, tables of fit results compared to dark matter models, and selected figures are presented in Appendix~\ref{data}, while individual galaxy case studies are highlighted in Appendix~\ref{results:MtoL}.

\section{  Rotation Curve Fitting Models}
 \label{sec:dos}

\subsection{Dark Matter Rotation Curve Formula}

The standard  rotation curve (RC) formula 
\begin{linenomath}
\begin{equation}
v(r)^2_{dyn}  =  v(r)^2_{lum}  +  v(r)^2_{dm}    
\label{eq:zonte1}
\end{equation}
\end{linenomath}
gives the dark matter model prediction   $v(r)_{dyn}$,     which is  fitted  to the rotation curve velocity parameter  $v(r)_{obs}$ which is a kinematic representation of  the observed
 Doppler-shifted spectra from a Lorentz boost.
 Terms in  $v(r)_{lum}$ represent the 
   orbital velocities  due to    the  luminous mass, interpreted by   classical Poisson gravity. Terms in $v(r)_{dm}$ represent the orbital velocities due    to the dark matter halo.

    \subsection{Relative-Frame Rotation Curve Formula} 

   We construct our model’s rotation curve formula by representing the standard dark matter contribution in Eq.~\ref{eq:zonte1} via a convolution of frame effects parameterized solely by luminous mass estimates. All transformations are defined one-to-one in radius between the observed galaxy and the host galaxy. Our modified  rotation curve formula is then  
\begin{linenomath} 
\begin{equation}
v(r)^2_{rc}  =  v(r)^2_{lum}  +   \alpha \kappa(r)^2 S(r)_{1} S(r)_{2},   
\label{eq:el2dm}
\end{equation} 
 \end{linenomath}  
 for   $v(r)_{rc}$   the model's prediction  which is  fitted  to the rotation curve velocity parameter  $v(r)_{obs}$.  We maintain the physical meaning of the  terms  $v(r)_{lum}$   as  representative of the   physical orbital  motion due to the luminous mass.
    We assume luminosity to be Lorentz scalar under a reliable distance indicator and therefore invariant to the relativistic transformations of 4-vectors below.  We note the  SPARC sample spans a distance range of $0.97$ to $125.65$ Mpc.  The term   $\alpha$ is  the model's free fitting parameter.

Terms in $\kappa(r)$  
are  ratios of the    Newtonian gravitational potentials from luminous mass estimates
\begin{linenomath}
\begin{equation}
\kappa(r)=\frac{\Phi(r)_{gal}}{\Phi(r)_{mw}} 
\label{eq:kappa2}   
\end{equation}
\end{linenomath}
for the galaxy being observed $\Phi(r)_{gal}$, by that of   the Milky Way $\Phi(r)_{mw}$.  
 Then terms   $S(r)_1$  are  defined by the transformation
\begin{linenomath}
 \begin{equation}
       S(r)_1 = \sinh \zeta(r), 
       \label{eq:hyperbolica}
   \end{equation}
   \end{linenomath}
  for the pseudo-rapidity $\zeta$ defined by  
  \begin{linenomath}
    \begin{equation}
     e^{\zeta(r)}=  \sqrt{\frac{(1-2\Phi(r)_{gal} ) }{(1-2\Phi(r)_{mw}) }}, 
      \label{eq:gravRS}
    \end{equation}
   \end{linenomath} 
   which represent the curved 2-frame mapping between the galaxies, where all terms are  ratios of  the Schwarzschild
clock terms in the weak field limit. Equivalently, terms can be seen as ratios    of the norms of the timelike Killing fields, which are invariants of the spacetime \citep{Wald}. 

 To use the Cartan frame-field technique, one is   required to transform the curved 2-frame back to the flat tangent frames where observations are made. 
We accomplish that transformation with 
  \begin{linenomath}
\begin{equation}
S(r)_{2} =  \cosh \tau(r)  
\label{eq:hyperbolico}
\end{equation}
 \end{linenomath}  
 for  the pseudo-rapidity $\tau$ defined by the sum of the pseudo-rapidity for the curved 2-frame transformation in Eq.~ \ref{eq:gravRS} and the pseudo-rapidity for  the flat frame-fields $\eta$, 
 \begin{linenomath} 
\begin{equation}
    e^{\tau(r)}=   e^{(\zeta(r)+\eta(r))}.
\end{equation}
 \end{linenomath} 
 Terms in  $e^{\eta(r)}$ return the transformations to the flat frame-fields
 \begin{linenomath} 
\begin{equation}
    e^{\eta(r)}=  \sqrt{\frac{1+\beta(r)}{1-\beta(r)}} 
    \label{eq:flat}
\end{equation}  
 \end{linenomath}  
 as they are defined  by the Keplerian velocity due to  the luminous mass
 $\beta(r) = v(r)_{lum}/c$, and are therefore our    best approximation of the flat frame-fields. This assumption is validated by the fact that dark matter is not required to reproduce the rotation curve of our solar system.
 
\begin{figure}[!ht]
\begin{adjustwidth}{-\extralength}{0cm}
\centering
\subfloat[\centering]{\includegraphics[width=13.0cm]{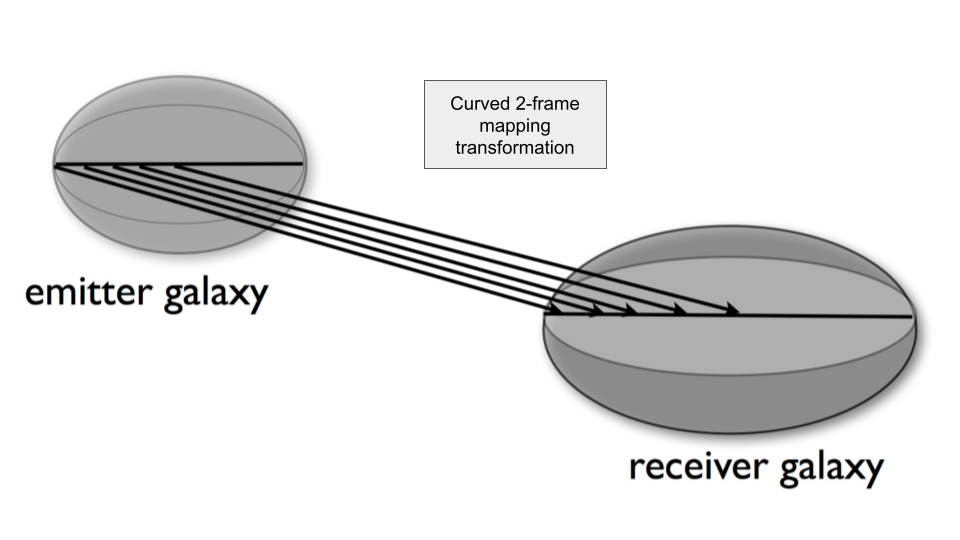}}\\
\subfloat[\centering]{\includegraphics[width=13.0cm]{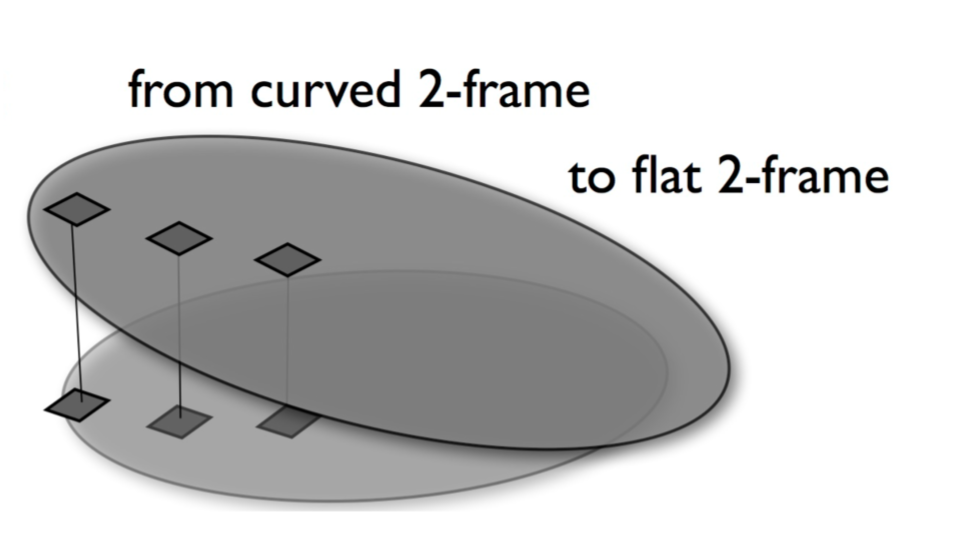}}
\end{adjustwidth}
\caption{   Graphical Representations of Our Transformations (\textbf{a}) Curved 2-frame transformation $S(r)_1$  and (\textbf{b}) curved-to-flat transformation $S(r)_2$}    \label{RCFMtransforms1and2} 
\end{figure}

 We emphasize that all terms on the right-hand side of our rotation curve formula, Eq.~\ref{eq:el2dm}, are now parameterized solely in terms of luminous mass and a single free parameter, $\alpha$. 
   Graphical representations of the two transformations are provided in Fig.~\ref{RCFMtransforms1and2}. 
   While our model is framed in the spirit of the Cartan field frames \cite{universe5100206},   we remain agnostic regarding the underlying physics of dark matter, and instead present a compact reformulation of the dark matter contribution that achieves good fits with fewer free parameters.
This heuristic treatment leverages the separability of effects on the frequency of light due to relative translational motion versus acceleration of the source \cite{Jack}, as well as the symmetry of the rotation curve data \cite{salucci}.

   Additionally, since Lorentz-type transformations are strictly applicable to inertial frames, we note \citet{rindler2013essential}'s observation that to compare galaxies as inertial frames is to compare them from    their respective centers. In Sec.~\ref{BC}, we demonstrate a suitable choice of boundary conditions for calculating the gravitational potentials, ensuring that galaxies are compared consistently from their respective centers.
   
\section{  Methods   } \label{sec:methods}

 \subsection{Luminous Mass Modeling } 
\label{lumAdd}

The baryonic mass components of a spiral  galaxy are represented  kinematically by  $v(r)_{disk}^2$ for the stellar disk, $v(r)_{bulge}^2$ for the stellar bulge, and $v(r)_{gas}^2$ for the gas halo,    
\begin{equation}
v(r)_{lum}^2 = \gamma_b v(r)_{bulge}^2 +  \gamma_d v(r)_{disk}^2 + v(r)_{gas}^2.  
\label{eq:zonte3}
\end{equation} 
Terms in $v(r)_{lum}$   represent the total expected circular orbital velocities about the galactic rotation axis due to the luminous mass. Terms in  $\gamma_b$ and $\gamma_d$ are    mass-to-light ratios  which  come from the complex process of modeling the  luminous mass associated with a given  spectral energy density as described in Sec.~\ref{intro}.
  Contributions from the gas   $v(r)_{gas}$  are calculated from a different observational technique  \citep{1983MNRAS.203..735C}, and do  not require  mass-to-light ratios.

  The only inputs to our model are  the Newtonian mass models published kinematically in the SPARC database   at \url{https://astroweb.case.edu/SPARC/MassModels_Lelli2016c.mrt} which come from the best photometry currently available. Detailed descriptions of the photometry and mass models are given in Appendix.~\ref{data}.
  

  \subsection{ Gravitational Potential \label{BC}}
    
 Classically the Newtonian gravitational potential  $\Phi(r)$ is   calculated by  the   integral
\begin{equation}
      \Phi(r) = - \int_{r_1}^{r_2} \vec{F} \cdot \vec{dr},  
      \label{eq:Newt}
\end{equation}
for $\vec{F}$ the central gravitational force and $\vec{dr}$ the path of the integration from an initial  value of $\Phi(r)_{lum} = 0$ at $r_1\to\infty$ to a negative maxima $\Phi(r)_{total}$, at the small limit $r_2\to 0$. 
In practice this integration     begins at   the largest observed radius $r_1=R_{max}$  and is summed  into a negative maxima at  the smallest observed radius  $r_2=r_{min}$. 
This classical   boundary condition is an implicit assumption of an asymptotically   flat embedding space, where the total energy goes to zero at infinity.  However   the    complexity of streamlines in the cosmic web 
  \citep{Pomarede:2020pme} demonstrate that  asymptotic flat space has not been recovered at the large 
  $R$ limit of observed galaxies.

Motivated by the astute observation of  \citet{rindler2013essential}, that    to compare
  galaxies inertially is to compare them from their respective centers, we   flip the bounds of integration such that all galaxies are compared from    a  common zero at $r\approx 0$ and summed to a positive definite value which approaches a constant for most galaxies. While the central black holes of galaxies do not have a mass of zero, their mass is vanishingly small with respect to the mass of the galaxy disk and bulge.
  

Gravitational potentials calculated in this way  satisfy the Poisson equation and populate Schwarzschild  clock terms $g(r)_{tt}=-(1-2\Phi(r))$. Since what we measure in practice is the difference in potentials, this technique allows galaxies to be compared inertially so that we can use Lorentz-type transformations in  Eq.~\ref{eq:hyperbolica} and \ref{eq:hyperbolico}, while maintaining contact with the classical usages of the potential.

\subsection{Geometric simplifications \label{GeomSphere}}

  The Luminous mass modeling problem for spiral galaxies (see Eq.~\ref{eq:zonte3})  usually assumes spherical symmetry for the 
  stellar bulge and gas halo, but   axial symmetry for    the stellar disk \citep{1954AJ.....59..273S,Freeman,Binney,2008MNRAS.391.1373B}.
 However, 
 it is a common technique to use   spherical symmetry for the entire mass distribution as proof of concept \citep{PhysRevDBekenstein2004, McGaugh_2008, 2022A&A...664A..40M,Loebman_2014}.
 Numerical integration of the thin stellar disk is  computationally intensive,  requiring  extra assumptions of under-constrained boundary conditions and  relevant physical scales\citep{2011A&A...531A..36H}.
For these reasons   we have   used the spherically symmetric approximation  and the Schwarzschild metric in this work. 
     
Gravitational potentials  calculated for  a spherical symmetry versus  a thin disk geometry converge to the same value  at lengths greater than one-third of the galaxy's exponential scale-length   $R_e$   \citep{Chatterjee}.  
  Since this is the region where   dark matter effects become important \citep{1985ApJAlbada}, this calculational technique is justified.
  However     in the inner region of a galaxy,  where our gravitational potential integrations begin,    
  the spherical assumption   
    overestimates the gravitational potential by a factor of $\approx  2$. This factor of $2$ can be clearly seen in our average fit results for the disk mass-to-light ratios   in Table~\ref{table:M2Light} and  \ref{TSet}.
       Implementation of an exponential   disk geometry would resolve this artifact of the analysis, but at significant computational expense \citep{1959HDP....53..275D}.

\subsection{Fitting procedure  }

A Python program was developed to fit galaxy rotation curve data using our model, and is publicly available at \url{https://github.com/Cisneros-Galaxy/RCFM}. The fitting procedure is as follows:  
first,  a       Milky Way baryon model is  selected.   Appendix~\ref{MWselect} compares the two Milky Way (MW) models used successively in this work. 
The MW model data in $v(r)_{lum}$ is  read into the program for a series of measurements in radii. 
  The galactic gravitational potential is then calculated by numerically integrating as in  Sec~\ref{BC}.   Once the MW potential is calculated, it remains static for the rest of the fitting procedure.  

Data for the   galaxies being observed include several pieces of information: the RC velocities $v(r)_{obs}$ from Doppler-shifted spectra, the uncertainty on that measurement $v(r)_{err}$, and the components of the luminous mass represented kinematically by    $v(r)_{bulge}$, $v(r)_{disk}$, and $v(r)_{gas}$ as  in Sec~\ref{lumAdd}. 
To calculate the baryonic potential for the galaxy in question, $v(r)_{lum}$ is first computed   as per Eq.~\ref{eq:zonte3}, with the starting values for the mass-to-light ratios of the bulge and disk respectively set at $\gamma_b=1$ and $\gamma_d=1$. Mass to light ratios are described in Section~\ref{lumAdd}.   The $\Phi(r)_{gal}$ associated with the galaxy   is then computed as in Sec~\ref{BC}.

After the potentials  due to  the  luminous mass of the galaxy being studied and the MW are  calculated,   the components must be compared at  matching values of $r$. To match radii, $\Phi(r)_{mw}$ is interpolated to produce values at the   radii reported in the observed galaxy
RC data $v(r)_{obs}$. Any galaxy observations with  radii larger than the largest radius in the MW model are discarded.

 The model's parameter $\alpha$ is set to its starting value, and the model prediction  is then   assembled  to give  $v(r)_{rc}$ 
 as in Eq.~\ref{eq:el2dm}, which is fitted to the galaxy's RC data $v(r)_{obs}$.  
 The parameters $\alpha$, $\gamma_b$ and $\gamma_d$ are allowed to freely vary,  recomputing $\Phi(r)_{gal}$ and $v(r)_{rc}$ at each step,  and optimal values are determined by the  minimization of the $\chi^2$.  The {\tt scipy.optimize.curve\_fit} utility in Python is used to perform this minimization.  This fit is repeated for several different starting values of $\alpha$ (powers of 10 from $10^{-3}$ to $10^{2}$) and the best fit is kept.  We fit the   175 galaxies in the SPARC database \cite{2016Lelli}, described in detail in Appendix~\ref{data}.

\subsection{Number of Free Parameters}

We present the model in this paper as a single-parameter framework. 
Here, we disregard the mass-to-light ratio parameters for the stellar disk and bulge, since the rotation curve fitting models we compare against (both dark matter and MONDian models) include these parameters in common. Dark matter models typically require between 2 and 9 free parameters, while MOND models require two, in addition to the stellar mass-to-light ratios.
Since the mass-to-light ratios are shared across dark matter and MOND models \cite{Milgrom,mcgaugh1998testing,McGaugh2016RAR}, we treat them as observational constraints from photometry and population synthesis modeling rather than as “free” parameters.

The under-constrained nature of luminous mass modeling arises because Doppler-shifted spectra, expected to constrain photometric measurements and population synthesis modeling, instead introduce additional free parameters in the context of dark matter halos. A class of models known as Forward Newtonian models, which predict baryonic mass densities directly from Doppler-shifted spectra, are not compared to this work; while these models represent a long-term goal, current formulations do not directly constrain luminous mass modeling.

With this accounting, dark matter models have a minimum of two free parameters \cite{JNav}, MONDian models have a minimum of two \cite{Bot}, and the present model has one.  Furthermore, as shown in Sec.~\ref{FreeCorrel}, the model’s free parameter appears highly correlated with a ratio of photometric quantities — luminosity and half-light radius — suggesting the possibility of a zero-parameter representation of dark matter phenomenology purely in terms of luminous mass estimates.

  \section{  Results  \label{sec:analysis} }  
\subsection{Goodness of Fits}

Rotation curve fitting models can be compared using two metrics obtained from the fitting procedure: the reduced $\chi^2_{r}$ values and the mass-to-light ratios. Since error estimates on rotation curve (RC) velocities have not been standardized across the field \citep{Blok1,Gent}, average $\chi^2_{r}$ values can only be meaningfully compared when fitted to the same RC data.

We fit the $175$ galaxy RCs in the SPARC database \cite{2016Lelli}, which have previously been analyzed with the Navarro-Frenk-White (NFW) dark matter model (using $\lambda$CDM priors) from \citet{Li_2020}. In the standard dark matter approach, the gravitational contributions from stars and gas are estimated, and the cold dark matter halo is parametrically modeled to predict the rotation curve velocity, which is then fitted to the observed Doppler-shifted velocities. For the entire SPARC sample, the NFW model has an average $\chi^2_{r} = 4.19$, while our model, using the Xue-Sofue-Jiao MW parameters \cite{Xue,Sofue,jiao2023detection}, has an average $\chi^2_{r} = 2.25$, and with the McGaugh-Jiao MW \cite{McGaugh_2019,jiao2023detection}, the average $\chi^2_{r} = 4.36$.

For the SPARC sample, the NFW dark matter model average mass-to-light ratios are $\gamma_{disk} = 0.51$ and $\gamma_{bulge} = 0.62$. For our model using the Xue-Sofue-Jiao MW, the averages are $\gamma_{disk} = 1.11$ and $\gamma_{bulge} = 0.77$, while for the McGaugh-Jiao MW, they are $\gamma_{disk} = 1.02$ and $\gamma_{bulge} = 0.68$. As expected, our disk mass-to-light ratios are approximately a factor of 2 larger than those from the NFW model due to the use of a standard spherical approximation for the galactic disk, which reduces computational complexity and minimizes excess parameters. As noted by \citet{Chatterjee}, the spherical approximation overestimates $\gamma_{disk}$ by a factor of $\approx 2$ at radii less than $R/3$, which is where our fits set the zero for the sums of the galaxy potentials. This implies that the true disk mass-to-light ratio for our model is closer to $\gamma_{disk} \approx 0.5-0.6$. These values are consistent with reported errors on luminous mass estimates of $\pm 20\%$ \cite{Lelli_2016surface}, and future developments will employ a full thin-disk geometry to improve accuracy.

 Taken together, these metrics indicate that our model provides a compact reformulation of the rotation curve fitting problem: it achieves comparable or better fits than the NFW model while using fewer free parameters. Individual galaxy results are reported in Table~\ref{table:M2Light}, for our model with respect to  the Xue-Sofue-Jiao MW alongside the NFW model for comparison.

\subsection{Free-parameter   Correlation  }
\label{FreeCorrel}

To frame a more reliable comparison, we select a subset of SPARC galaxies with the highest quality data based on the following criteria:
\begin{enumerate}
\item Galaxies with the most accurate distance estimates (Tip of the Red Giant Branch \citep{McQuinn_2019} and  Cepheid variable stars \citep{10.1046/j.1365-8711.2003.06786.x}), rejecting all others. \
\item  Galaxies with sky inclinations between $15^\circ$ and $80^\circ$, excluding those with inclinations greater than $80^\circ$, where surface brightness profiles are uncertain, and less than $15^\circ$, where line-of-sight Doppler shifts are unreliable. \
\item Excluding galaxies with a quality factor $Q = 3$, which are not suited for dynamical studies. The SPARC database \citep{2016Lelli} assigns $Q = 1$ for high-quality rotation curves, $Q = 2$ for minor asymmetries or lower-quality curves, and $Q = 3$ for major asymmetries, strong non-circular motions, or offsets between HI and stellar distributions.
\end{enumerate}

This selection results in a subset of 36 galaxies, reported in Table~\ref{TSet}, representing the highest-quality rotation curves and photometric models with well-established distances. All galaxies are local ($<15$ Mpc) and undisturbed, which minimizes distance-related selection biases and includes a variety of dwarf, gas-dominated galaxies.

For this subset, the NFW dark matter model has an average $\chi^2_{r} = 5.46$ and fails to fit two galaxies (NGC6789 and UGC07232). Our model fits all galaxies, yielding an average $\chi^2_{r} = 1.64$ for the Xue-Sofue-Jiao MW \cite{Sofue,Xue,jiao2023detection} and $\chi^2_{r} = 1.65$ for the McGaugh-Jiao MW \cite{McGaugh_2019,jiao2023detection}.

The two MW models differ primarily in central concentration: the Xue-Sofue-Jiao MW has a de Vaucouleurs stellar bulge, whereas the McGaugh-Jiao MW has a more diffuse bulge-bar structure. Our model, tested on the full SPARC sample, slightly prefers the Xue-Sofue-Jiao MW, but for the high-quality subset, both MW models yield remarkably low $\chi^2_{r}$ values, providing confidence in the model’s consistency with the SPARC rotation curve data. At this level, there is no strong distinguishing feature between MW models.

\begin{figure}[!ht] 
\begin{adjustwidth}{-\extralength}{0cm}
\centering 
\subfloat[\centering]{\includegraphics[width=7.0cm]{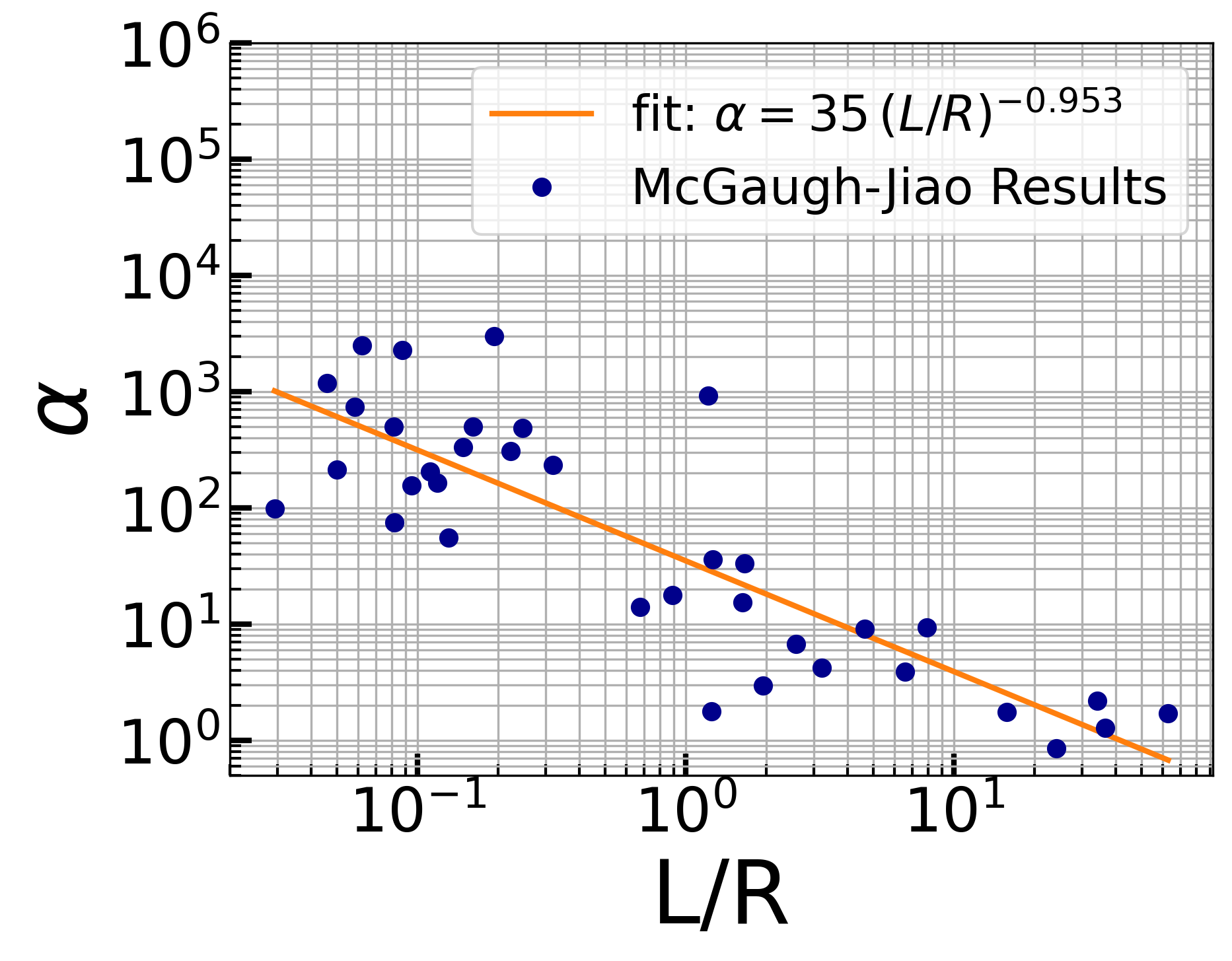}}\\
\subfloat[\centering]{\includegraphics[width=7.0cm]{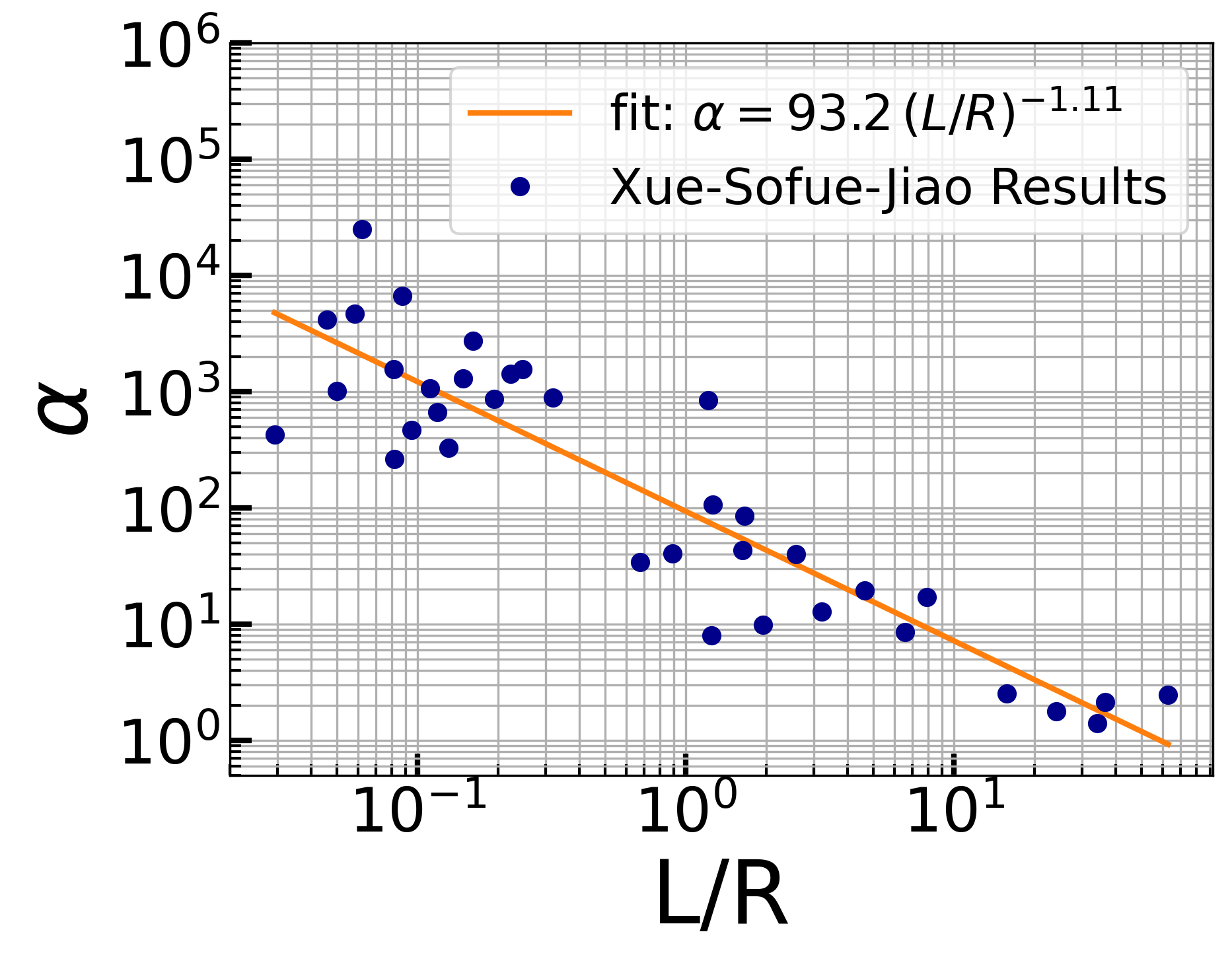}}
\end{adjustwidth} 
\caption{\textbf{  Physical Interpretation of the Model's free parameter $\alpha \propto L/R$ }\\ 
  For $L$ the total luminosity in units of $10^9 M_{\odot}$ and $R$
the   half-light radius   in units of  $kpc$.   There is one $\alpha$   value per galaxy  in the subset of the most reliable  SPARC galaxy data described in     Table~\ref{TSet} and    Sec.~\ref{sec:analysis}. We report results with respect to  the    MW baryon model from (\textbf{a}) McGaugh-Jiao and (\textbf{b}) Xue-Sofue-Jiao.  \label{alpha2} }
\end{figure}

To explore the physical interpretation of the model’s free parameter, $\alpha$ (Eq.~\ref{eq:el2dm}), we plot the ratio of total luminosity $L$ to half-light radius $R$ from   an auxiliary file in the SPARC database \cite{2016Lelli} versus the fitted $\alpha$ for each galaxy in the subset. Here, $L$ is in units of $10^9 L_\odot$ and $R$ in kpc, both measured in the 3.6 $\mu$m band with a solar absolute magnitude of $3.24$ \cite{oh2008high}. The resulting distributions are well fit by power laws (Fig.~\ref{alpha2}):

For the Xue-Sofue-Jiao MW, we  find the $\alpha$ parameter form  is 
\begin{equation}
   \alpha = 93.2 \left(\frac{L}{R}\right)^{-1.11}
   \label{XSJmw}
\end{equation}
with a coefficient of determination of $R^2 = 0.82$.
  For the McGaugh-Jiao MW, we  find the $\alpha$ parameter form  is
 \begin{equation}
   \alpha = 35 \left(\frac{L}{R}\right)^{-0.953}
   \label{MGmw}
\end{equation}
with a coefficient of determination of $R^2=0.72$.
These power-law fits are shown   in Fig.~\ref{alpha2}.

Since all other terms in Eq.~\ref{eq:el2dm} are ratios of galaxy quantities relative to the Milky Way, and $\alpha$ is dimensionless, we propose that $\alpha$ may itself be expressed as a ratio of $L/R$ terms: 
\begin{equation}
   \alpha =   \left(\frac{(L/R)_{mw}}{(L/R)_{gal}}\right)^{b}
\end{equation}
Using the coefficients from Eqs.~\ref{XSJmw} and \ref{MGmw}, this correlation function predicts a MW luminosity given its half-light radius. For the Xue-Sofue-Jiao MW, this yields $L_{MW} = 0.89 \times 10^9 L_\odot$ with $R = 0.5$ kpc; for the McGaugh-Jiao MW, $L_{MW} = 6.85 \times 10^9 L_\odot$ with $R = 4.2$ kpc. As we are embedded in the Milky Way, it is difficult to determine which estimate is more physically realistic, though upcoming surveys such as the Vera C. Rubin Observatory’s LSST \cite{Ivezić_2019} will continue to refine constraints on the Milky Way’s structure and mass content.

 Taken together, the strong correlation between $\alpha$ and photometric quantities suggests that, for high-quality data, the model’s single free parameter may be largely predictable from luminous mass estimates alone. In other words, this provides a methodological pathway toward a zero-parameter representation of rotation curve fits, without making claims about the existence or absence of dark matter.

   \subsection{Comparing Milky Way models}

The model presented in this paper requires a static background choice for the Milky Way (MW) baryon distribution.  Determining the gravitational potential of the MW from our position inside the disk is notoriously difficult \cite{10.1093/mnras/stt814,1991ARA&A..29..409F}, and the galaxy’s baryon distribution remains an active area of research. Previous studies \cite{Iocco_2015} suggest that a substantial dark matter component is required, while Jeans equation analyses indicate that long-term dynamical stability of a disk necessitates a large shell of concentric matter \cite{1973ApJ...186..467O,Loebman_2014}. More recent analyses of GAIA DR3 data \cite{Koop_2024} suggest that the MW may require only a minimal dark matter halo, and \citet{jiao2023detection} indicate that the MW rotation curve is approximately Keplerian out to at least $26.5$ kpc.

 Regardless of the outcome of this debate, our goal is not to weigh in on the existence of MW dark matter halos. Instead, we introduce a simple, novel reformulation of the contribution commonly attributed to dark matter. By encoding these effects as frame-field transformations parameterized by luminous mass estimates, the problem becomes more tractable, with only a single free parameter.

 Due to ongoing uncertainties regarding the MW, we have chosen two widely used MW baryon models to fit the SPARC galaxy sample. A detailed summary of these two MW models is provided in Sec.~\ref{MWselect}, and a comparison of the results is given in Table~\ref{TSet}. The main distinction between the models lies in their central mass concentration and the framework used to extend the   rotation curves due to the baryons.

The MW model from Xue-Sofue \cite{sofue2009unified,Xue} has a highly concentrated central mass, represented by a de Vaucouleurs-law stellar bulge with a half-mass scale radius of $0.5$ kpc.  It is modeled using the NFW dark matter framework combined with Jeans Equation analysis informed by SDSS observations of 2,400 blue horizontal branch stars.

The MW model from McGaugh \cite{McGaugh_2019} has a more diffuse triaxial bulge-bar with a scale length of $4.2$ kpc.  It is modeled using the MOND-RAR framework and Jeans Equation analysis informed by GAIA and APOGEE data. Based on average $\chi^2_{r}$ values, our model slightly favors the centrally condensed Xue-Sofue MW over the McGaugh MW. However, for the subset of galaxies with the most reliable data, the two MW models produce virtually indistinguishable fits.  In future work, we plan to extend our analysis to all 56 MW baryon models from \citet{2019MNRAS.487.5679L}, exploring the impact of different bulge, disk, and gas halo assumptions on high-quality rotation curves.

\begin{figure}[!ht] 
\begin{adjustwidth}{-\extralength}{0cm}
\centering 
\subfloat[\centering]{\includegraphics[width=7.0cm]{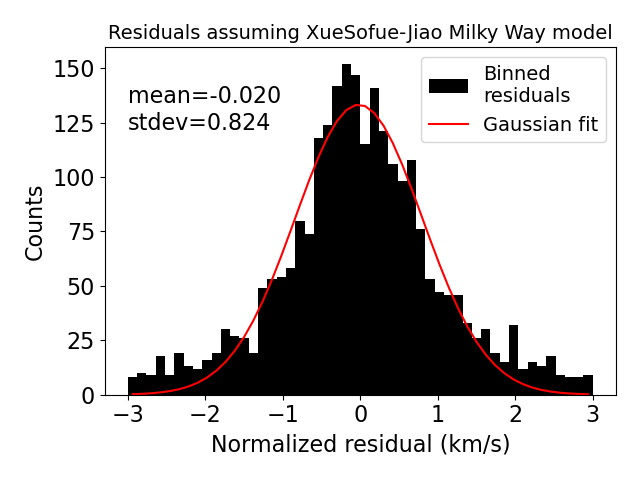}}
\subfloat[\centering]{\includegraphics[width=7.0cm]{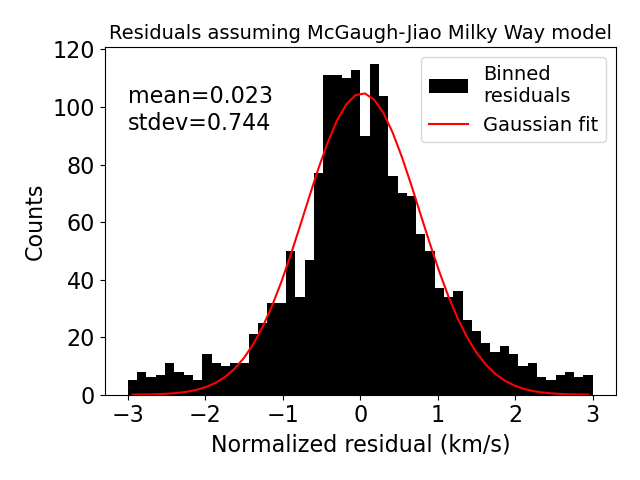}}\\
\end{adjustwidth} 
\caption{Normalized residuals for fits to SPARC galaxies assuming two different MW baseline models, each   fitted by a Gaussian  function. The means and standard deviations of the Gaussian fit are shown.   \label{fig1} 
(\textbf{a}) For the Xue-Sofue-Jiao Milky Way (\textbf{b}) For the McGaugh-Jiao Milky Way }
\end{figure}

Residuals for each MW model were computed by subtracting the model-predicted velocities from the observed SPARC rotation velocities at each radius. Figure~\ref{fig1} shows the distributions of these normalized residuals for all 175 galaxies and both MW models. In all cases, the residuals are narrowly centered around zero with a range of approximately $\pm 3$ standard deviations, though with some heavy tails. Their similar behavior across MW models indicates that our fitting parameters are robust to the choice of MW baryon distribution.

Gaussian fits to the residual distributions, obtained using {\tt scipy.curve\_fit} function in Python, yield the following: for the McGaugh-Jiao MW model, mean $= 0.023$ and standard deviation $= 0.744$; for the Xue-Sofue-Jiao MW model, mean $= -0.020$ and standard deviation $= 0.824$ (all in units of standard deviations in velocity error). The small values associated with these quantities in both cases provide confidence that our fits match data closely.

   \subsection{ Comparisons to Other Alternative Formulations of the Rotation Curve Problem} 
\label{otros}

Alternative approaches have been proposed to explain the observed linkages between luminous and dark matter in galaxies. The most widely known is Modified Newtonian Dynamics (MOND) \cite{Milgrom}, which modifies classical gravity by introducing a characteristic acceleration scale $a_0$. MOND and the Radial Acceleration Relation (RAR) have been applied to all SPARC galaxies, yielding average $\chi^2$ values comparable to NFW dark matter fits. Both MOND and dark matter models assume that Doppler-shifted spectra directly reflect the  rotation velocities.

Another class of model are the  Newtonian orbital models (NOMs) which rely solely on luminous mass without modifying gravity. Examples include \citet{galaxies8020036} and \citet{galaxies8010009}, which solve the inverse problem of inferring baryonic mass directly from rotation curves. While promising as zero-parameter models, NOMs achieve full consistency with photometry in only $\sim 62\%$ of galaxies; the remainder require additional low-luminosity mass components. 
 
 Other alternative efforts look at geometry and the characterization of the gravitational potential, such as \citet{doi:10.1139/cjp-2016-0625}, who analyze the problem using oblate spheroid spin rather than thin-disk approximations.  As discussed  in Sec.~\ref{BC}, the thin-disk gravitational potential   differs from that of a spherical mass distribution at both small radius, where a disk-like geometry requires more mass for a given acceleration,  and at large radii where less mass is required \cite{Chatterjee}. So, by characterization of the oblate spheroid shape of real galaxies and spin, instead of simply a thin-disk, 
\citet{doi:10.1139/cjp-2016-0625}   find  a   model which   requires only $4/9$ of the baryonic mass predicted by NOMs. 
Another interesting approach from \citet{galaxies8010009} introduces a “galactic rotation parameter” $A$ to account for deviations from spherical symmetry, with $A$ ranging from $1$ to $2$.   However, some predicted densities, particularly for the MW, exceed observational constraints \cite{creze1997}.
These studies provide interesting insights into the problem, but are difficult to compare directly to our   model due to the framing of our results in terms of reduced $\chi^2$ and mass-to-light ratios.

Relativistic formulations, including TeVeS \cite{PhysRevDBekenstein2004} and the metric-skew-tensor model \cite{Brownstein_2006}, reproduce MOND successes for galaxy rotation curves without non-baryonic dark matter. TeVeS introduces vector and scalar fields to modify gravity but suffers from nonphysical caustic instabilities at high energies. Brownstein’s approach fits $\sim100$ galaxies but does not report $\chi^2$ or mass-to-light ratios, making direct comparison difficult. Importantly, both modify classical gravity, whereas our model retains standard relativistic physics and instead encodes dark matter-like effects through Cartan frame-field transformations of the host galaxy.

In summary, while MOND, NOMs, and relativistic modifications provide valuable insights, they either modify fundamental physics or remain incomplete. In contrast, our model fits all 175 SPARC galaxies with a single free parameter, using only luminous mass estimates and standard relativistic principles.
 
\section{Conclusions\label{sec:conclu}}

The early view of   dark matter   was  that the galactic \citep{1978Rubin} and cosmological \citep{2010dmp..book.....S,Tully:2014gfa,Naidu_2022}   problems were distinct. We follow that precedent here,   focusing exclusively on the rotation-curve problem.
 The model presented in this paper encodes the dark matter contribution in the standard rotation curve problem as a product of two relative galaxy frame transformations, parameterized solely by estimates of the luminous mass and a single free parameter. This result is then compared to dark matter models that typically employ at least two free parameters.

 These transformations are constructed in the spirit of the Cartan frame-fields, where all terms are ratios of the timelike Killing fields between the galaxy being observed and the Milky Way observing frame.  We avoid interpreting the physics of this approach with respect to the debate over the existence of dark matter, and instead simply demonstrate that encoding dark matter effects in terms of luminous mass provides excellent fits to a large sample of high-quality rotation curves. Independent of the physical interpretation of this model, it represents a one-parameter fit to rotation curve data, parameterized solely by estimates of the luminous mass, and constructed to reproduce the standard dark matter contribution to rotation curves.

We test our model on a sample of  175 high-quality rotation curves of spiral galaxies from the SPARC database \citep{2016Lelli} previously fitted by a NFW dark matter model,  with $\lambda$CDM priors, from  \citet{Li_2020}.
For the SPARC sample, our model with the Xue-Sofue-Jiao Milky Way  results in  an average reduced $\chi^2_{r}=2.24$, whereas the NFW dark matter model gives an average reduced $\chi^2_{r}=4.19$.

   This approach re-purposes traditional physics transformations, to compare galaxies one-to-one in radius, but does not modify physics.
Early investigations    into the use of relativity to resolve the galaxy rotation curve problem   used Galilean subtraction of the potentials at  large radii \citep{MTW}, versus  frame-field transformations   appropriate to the four-vector algebra of Doppler-shifted frequencies.  In addition, we show   how a possible  physical interpretation of the    model's free parameter in terms of  $L/R$, luminosity and effective radius, 
 may be possible. 
 Interpreting  our  model’s   free-parameter   with a ratio of two photometric  quantities could   lead to   a zero parameter model and a direct constraint to  luminous mass modeling  from the orthogonal observation of Doppler-shifted spectra. 
 The  upcoming observations   from the Vera C. Rubin Observatory's Legacy Survey of Space and Time  \citep{Ivezić_2019} will  better inform    the future Milky Way model used  in this work.

\authorcontributions{   
The work in this  paper   was divided as follow; conceptualization was by S.Cisneros;  
methodology was by S.Cisneros, R.Ott and A.Roberts;  
software, by R.Ott, A.Roberts, M.Crowley and M.Paz;  
validation,by R.Ott and A.Roberts;   
formal analysis was by S.Cisneros, M.Crowley, R.Ott, A.Roberts, and M.Paz;  
investigation was by S.Cisneros, M.Crowley  and M.Paz;  
data curation was S.Cisneros, M.Crowley, R.Ott, A.Roberts, and M.Paz; 
writing---original draft preparation was by  S.Cisneros;  
 writing---review and editing was by R.Ott, A.Roberts, and M.Crowley;  
visualization was by R.Ott, M.Crowley and M.Paz; and
supervision was by S.Cisneros.  
 All authors have read and agreed to the published version of the manuscript. }

\funding{This research received no external funding.}

\dataavailability{The code used to generate all figures and fit results is publicly available at \url{https://github.com/Cisneros-Galaxy/RCFM}. The rotation curve data and photometric models used in this paper are from the SPARC database publicly available at \url{http://astroweb.cwru.edu/SPARC/}. The few additional datasets included for comparison are available at the cited references.  }

\acknowledgments{ 
This work is dedicated to Emmett Till. 
 We acknowledge and express     gratitude to the first nations peoples for their generosity,  on whose unceded lands this was written; including but not limited  to  the  
 the Coos, Lower Umpqua and Siuslaw Indians of Oregon, the Coast Salish bands of the Puget Sound, 
 the Cheyenne and Arapaho Tribes and  Ute  Tribes of Colorado,  the Navajo Nation and Pueblos of New Mexico, and  the Algonquian and Iroquoian  Peoples of Massachusetts and New York. 
  The authors would like to thank    V.\,P.\,  Nair, T.\, Boyer,  M.\, Kaku, I.\, Chavel,  R.\, Walterbos, N.\,P.\, Vogt,
  S.\, McGaugh, A.\, Klypin, T.\, Quinn, S.\ Tuttle, M.\, Juric,  R.\, Rivera, N.S.\, Oblath,  J.\, Formaggio, J.\, Conrad,  P.\, Fisher, Y.\, Sofue, C.\, Mihos, and M.\, Merrifield for their guidance. We wish to also thank the many students who have contributed to this work over the years, including but not limited to; A.\, Ashley, D.\, Battaglia, D.\, Chester, R.\, Robinson, A.\, Rodriguez, Z.\, Brown, P.\, Pham, Z.\, Holland, A.\, Livingston, L.\, Castrellon, R.\, Real Rico, E.\, Gutierrez-Gutierrez, S.J.\, Rubin, S.\, Graham, and L.\, Joyal. We thank ChatGPT for helping to steamline language and to improve grammar and flow. Finally, we sincerely thank the reviewers and the editor for their time and careful attention, which greatly improved the clarity and overall quality of the manuscript. 
 }

\conflictsofinterest{ The authors declare no conflicts of interest.} 



\abbreviations{Abbreviations}{
The following abbreviations are used in this manuscript:\\
\noindent 
\begin{tabular}{@{}ll}
RC & Rotation Curve\\
MW & Milky Way\\
SPARC & Spitzer Photometry and Accurate Rotation Curves \\
URC & Universal Rotation Curve\\
SED & Spectral Energy Density\\
SDSS & Sloan Digital Sky Survey\\
$\lambda$CDM & $\lambda$ Cold Dark Matter\\
RAR & Radial Acceleration Relation\\
NFW & Navarro-Frenk-White\\
NOMs & Newtonian orbital models  \\
TeVeS & Tensor-Vector-Scalar  
\end{tabular}
}

\appendixtitles{yes} 
\appendixstart
\appendix

  \section[\appendixname~\thesection]{Figures, Tables and Detailed Information on Data used}
\label{data}
\subsection[\appendixname~\thesubsection]{SPARC Database}

   We fit the Spitzer Photometry and Accurate Rotation Curves (SPARC) dataset  of  175 nearby late type galaxies with the model presented in this paper. The SPARC sample provides   extended RC data from atomic hydrogen (HI)  and H-$\alpha$ \citep{2016Lelli}. 
HI  provides the most reliable
 RCs because it is dynamically cold, traces circular orbits, and can be observed several effective radii past the stellar disk. 
 This sample of rotationally supported galaxies   spans the widest range of masses and morphologies presently available. 
 
These galaxies are  reported with    Newtonian luminous mass models, represented kinematically  as in Eq.~\ref{eq:zonte3}. These baryon mass distributions are    based on 
   Spitzer Photometry in the 
   near infrared  at 3.6$\mu m$.
   Near infrared  is    believed to be the best tracer of stellar mass \citep{10.1093/mnras/sty3223}, as at this wavelength     mass-to-light ($\gamma_i$) ratios   are believed to be almost constant and  independent of star formation history \citep{BelldYong,10.1093/mnras/sty3223}. 
  The SPARC database  reports     mass-to-light ratios set to  $\gamma_i=1$ in units of $M_{\odot} / L_{\odot}$   at 3.6$\mu m$ for all galaxies.  
 
 SPARC  gas fractions are reported as $v(r)_{gas}$,  calculated from surface density profiles of HI   with the formalism given in  \citep{1983MNRAS.203..735C} and  scaled by
     a factor 1.33 to account for cosmological helium abundances.     The addition of molecular gas could increase mass fractions in the inner kiloparsec of a galaxy   \citep{2004ApJ...609..652M},  but are not included as most galaxies in the SPARC sample do not have  molecular measurements available.

     Error on these velocities is estimated at $\pm 20\%$ \citep{2016Lelli}. The      SPARC  database can be found  at \url{http://astroweb.cwru.edu/SPARC/}.

\subsection[\appendixname~\thesubsection]{Milky Way Luminous Mass Models}
\label{MWselect}

  Our model  requires a static choice of a MW  luminous mass   model to fit our observations of external galaxies.  In this paper, we   compare the SPARC sample of 175 galaxies successively to
  two different MW  baryon models.

  The MW model from   \citet{McGaugh_2019} has a triaxial bulge-bar in the center of the luminous mass distribution and the extended rotation curve comes from Radial Acceleration Relation (RAR) \cite{Lelli_2017},  a phenomenological model arising from analysis of  Modified Newtonian Dynamics \cite{Milgrom}. 
  The MW from  Sofue  and Xue \citep{Sofue,Xue}   has a de Vaucouleurs bulge and is extended by NFW dark matter modeling \cite{NFW}. 
  The difference between the  two MW models is marked in the inner   $7$ kpc; as the McGaugh MW has a scale-length of $4.2$ kpc and the   Xue-Sofue MW model has a scale-length of $0.5$ kpc. 
 The Sofue model covers a range of    $0$ to $20$ kpc,   and   the Xue MW model from $20$ to $60$ kpc.  The Sofue and Xue MW models are   from the same data and   dark matter  model  extension, so are combined   to increase coverage.

   The  recent ESA \emph{Gaia}
   data release (DR3),  \cite{jiao2023detection} demonstrates   a Keplerian decline of the rotation curve   of the MW in the range of  $9.5$ to $26.5$ kpc, consistent with our model.      The ESA \emph{Gaia} mission, taking measurements from  the second Lagrange point,  has revolutionized the science of the   MW with
     unprecedented detail,  statistical accuracy,  and a drastic reduction in  systematic uncertainties.   Since this is the both the most recent and best data for the MW, we  replace  both models' MW   velocities in the range from $9.5$ to $26.5$ kpc with those from    \cite{jiao2023detection}. We  then shift the two   sides of the given MW  model     by a constant amount  to match   the endpoints of the Jiao rotation curve. Hence, the Xue-Sofue MW model becomes the Xue-Sofue-Jiao MW and the McGaugh MW becomes the McGaugh-Jiao MW. 
     See Table \ref{MW_dats} for summary details.

\begin{table}[h!] 
\caption{ Milky Way Models\\
\emph{NOTE:}  The table lists the original model sources and assumptions. For both Milky Way models, we have replaced the original velocity data in the region spanning $9.5$–$26.5$ kpc with the Gaia DR3 velocities reported by \citet{jiao2023detection} due to the high quality of the observations. Throughout the paper, these modified models are referred to as Sofue–Xue–Jiao and McGaugh–Jiao.  \label{MW_dats} }
\begin{tabularx}{\textwidth}{CCCC}
\toprule
\textbf{Author}	&  \textbf{scale length } 	& \textbf{ Model  for   } &\textbf{Range}\\
                  &	   bulge/bar           &	extended RC      &	\\
\midrule
Sofue-Xue  	            &0.5 kpc     &	NFW dark matter  	&[0,60] kpc \\
\cite{Sofue,Xue}        &  bulge    &     \citep{NFW}         &	 	\\
\hline				
McGaugh                 & 4.2 kpc   &	RAR   	          &[0,150] kpc \\
\cite{McGaugh_2019}     &bulge-bar        &   \citep{Lelli_2017}    &	\\
\bottomrule
\end{tabularx} 
\end{table}

\subsection[\appendixname~\thesection]{  Fit Results for  the   175 SPARC galaxies}
\label{FIGyTAB}

 This paper's model results in Table~\ref{table:M2Light} are reported with respect to the Xue–Sofue–Jiao Milky Way model  \cite{Xue, sofue2009unified,jiao2023detection}, because it performs significantly better than   the  McGaugh-Jiao Milky Way for the whole sample - though the two Milky Way models perform equally well on a more reliable subset of the SPARC galaxies. Table~\ref{TSet} reports our results for the  subset for both Milky Way models,  the McGaugh-Jiao   and the Xue-Sofue-Jiao.  Stellar mass-to-light ratios, $\gamma_i$, are given in units of solar mass per solar luminosity ($M_\odot/L_\odot$).  
 Galaxies for which dark matter model fits fail are indicated by NaN. 

\setlength\LTleft{0pt}
\setlength\LTright{0pt}
 \begin{center} 
\begin{longtable}{|l|ccc|cccr|}
\caption{ Fit Results for  the SPARC 175 galaxy   sample \label{table:M2Light}\\
 Luminous   models which did not include bulge components as reported in the SPARC database of Newtonian mass models are indicated by three dots.}   \\
\hline
\multicolumn{1}{|c|}{Name } & \multicolumn{3}{c|}{\textbf{NFW Dark Matter}} & \multicolumn{4}{c|}{\textbf{Our Model with XSJ MW }}   \\  
 \multicolumn{1}{|c|}{\textbf{ }} & \multicolumn{1}{c}{$\chi^2_{r}$ } & \multicolumn{1}{c}{$\gamma_{disk}$ } & \multicolumn{1}{c|}{$\gamma_{bulge}$ } 
 & \multicolumn{1}{|c}{$\chi^2_{r}$ } & \multicolumn{1}{c}{$\gamma_{disk}$ } & \multicolumn{1}{c}{$\gamma_{bulge}$ } & \multicolumn{1}{c|}{$\alpha$ }   \\ \hline             
\endfirsthead
\multicolumn{3}{c}%
{{\bfseries \tablename\ \thetable{} -- continued from previous page}} \\
\hline 
\multicolumn{1}{|c|}{Name} & \multicolumn{3}{c|}{\textbf{NFW Dark Matter}} & \multicolumn{4}{|c}{\textbf{ Our Model with XSJ MW}}   \\  
 \multicolumn{1}{c}{\textbf{ }} & \multicolumn{1}{|c}{$\chi^2_{r}$ } & \multicolumn{1}{c}{$\gamma_{disk}$ } & \multicolumn{1}{c|}{$\gamma_{bulge}$ } 
 & \multicolumn{1}{|c}{$\chi^2_{r}$ } & \multicolumn{1}{c}{$\gamma_{disk}$ } & \multicolumn{1}{c}{$\gamma_{bulge}$ }  & \multicolumn{1}{c|}{$\alpha$ }  \\ \hline  
\endhead
\hline \multicolumn{8}{|r|}{{Continued on next page}} \\ \hline
\endfoot
\hline 
\endlastfoot
\hline
CamB & 7.92 & 0.35 & … & 0.23 & 0.00 & … & 24,989.21 \\
D512-2 & NaN & NaN & NaN & 0.21 & 1.48 & … & 131.01 \\
D564-8 & 6.7 & 0.45 & … & 0.11 & 1.21 & … & 4,130.26 \\
D631-7 & 12.21 & 0.39 & … & 0.30 & 0.26 & … & 2,725.54 \\
DDO064 & 1.05 & 0.5 & … & 0.45 & 1.58 & … & 345.48 \\
DDO154 & 16.57 & 0.33 & … & 10.58 & 1.18 & … & 1,567.39 \\
DDO161 & 1.99 & 0.47 & … & 0.62 & 0.97 & … & 279.60 \\
DDO168 & 23.93 & 0.56 & … & 4.36 & 0.76 & … & 1,290.43 \\
DDO170 & 4.44 & 0.5 & … & 3.19 & 1.84 & … & 122.03 \\
ESO079-G014 & 4.9 & 0.57 & … & 3.86 & 1.09 & … & 8.46 \\
ESO116-G012 & 3.82 & 0.39 & … & 1.03 & 1.04 & … & 58.84 \\
ESO444-G084 & 4.71 & 0.52 & … & 0.40 & 1.86 & … & 465.01 \\
ESO563-G021 & 19.23 & 0.96 & … & 16.41 & 1.00 & … & 1.78 \\
F561-1 & 1.21 & 0.49 & … & 1.05 & 0.96 & … & 0.00 \\
F563-1 & 1.47 & 0.54 & … & 0.95 & 2.06 & … & 60.28 \\
F563-V1 & 2.21 & 0.47 & … & 0.29 & 0.99 & … & 0.00 \\
F563-V2 & 1.98 & 0.55 & … & 0.11 & 2.20 & … & 11.90 \\
F565-V2 & 3.6 & 0.54 & … & 0.31 & 2.22 & … & 218.59 \\
F567-2 & NaN & NaN & NaN & 0.50 & 1.31 & … & 29.40 \\
F568-1 & 1.57 & 0.56 & … & 0.72 & 1.91 & … & 27.98 \\
F568-3 & 4.99 & 0.55 & … & 1.79 & 1.31 & … & 68.81 \\
F568-V1 & 0.31 & 0.53 & … & 0.14 & 2.14 & … & 9.17 \\
F571-8 & 9.6 & 0.2 & … & 2.04 & 0.17 & … & 12,596.96 \\
F571-V1 & 1.39 & 0.5 & … & 0.20 & 1.49 & … & 102.53 \\
F574-1 & 2.21 & 0.54 & … & 1.42 & 1.54 & … & 20.27 \\
F574-2 & NaN & NaN & NaN & 0.14 & 0.67 & … & 147.79 \\
F579-V1 & 0.91 & 0.47 & … & 1.07 & 1.63 & … & 0.00 \\
F583-1 & 2.23 & 0.59 & … & 1.04 & 1.87 & … & 128.01 \\
F583-4 & 0.34 & 0.49 & … & 0.28 & 1.30 & … & 60.22 \\
IC2574 & 36.3 & 0.44 & … & 2.27 & 1.10 & … & 885.47 \\
IC4202 & 20.56 & 0.67 & 0.14 & 12.18 & 0.00 & 0.44 & 150.83 \\
KK98-251 & 3.61 & 0.53 & … & 0.42 & 1.67 & … & 536.07 \\
NGC0024 & 0.66 & 1.18 & … & 0.73 & 1.39 & … & 9.88 \\
NGC0055 & 4.35 & 0.36 & … & 2.86 & 1.01 & … & 106.90 \\
NGC0100 & 1.94 & 0.46 & … & 0.10 & 0.93 & … & 161.92 \\
NGC0247 & 2.03 & 0.66 & … & 2.18 & 1.53 & … & 8.03 \\
NGC0289 & 2.06 & 0.53 & … & 1.78 & 0.74 & … & 3.59 \\
NGC0300 & 0.93 & 0.41 & … & 0.42 & 1.14 & … & 85.28 \\
NGC0801 & 7.15 & 0.72 & … & 7.38 & 0.77 & … & 0.82 \\
NGC0891 & 4.39 & 0.25 & 0.54 & 1.89 & 0.66 & 0.00 & 3.46 \\
NGC1003 & 2.54 & 0.66 & … & 3.42 & 0.77 & … & 80.54 \\
NGC1090 & 2.77 & 0.54 & … & 2.27 & 0.81 & … & 4.88 \\
NGC1705 & 1.33 & 0.82 & … & 0.13 & 1.25 & … & 35.94 \\
NGC2366 & 3.91 & 0.31 & … & 2.34 & 1.06 & … & 446.13 \\
NGC2403 & 9.09 & 0.39 & … & 10.73 & 0.86 & … & 19.32 \\
NGC2683 & 2.66 & 0.66 & 0.67 & 1.01 & 0.88 & 0.44 & 1.78 \\
NGC2841 & 1.47 & 1 & 0.88 & 1.36 & 0.91 & 1.10 & 1.40 \\
NGC2903 & 6.54 & 0.27 & … & 7.45 & 0.62 & … & 2.62 \\
NGC2915 & 1.05 & 0.44 & … & 0.63 & 0.56 & … & 841.91 \\
NGC2955 & 4.27 & 0.32 & 0.82 & 4.45 & 0.40 & 0.88 & 1.80 \\
NGC2976 & 1.45 & 0.6 & … & 0.46 & 0.91 & … & 40.19 \\
NGC2998 & 3.53 & 0.55 & … & 3.74 & 0.87 & … & 2.14 \\
NGC3109 & 15.31 & 0.92 & … & 0.28 & 2.05 & … & 661.98 \\
NGC3198 & 1.46 & 0.46 & … & 1.72 & 0.88 & … & 8.51 \\
NGC3521 & 0.29 & 0.5 & … & 0.77 & 0.71 & … & 1.67 \\
NGC3726 & 3.27 & 0.49 & … & 2.41 & 0.76 & … & 8.32 \\
NGC3741 & 1.74 & 0.55 & … & 0.65 & 0.71 & … & 6,628.19 \\
NGC3769 & 1.06 & 0.41 & … & 0.71 & 0.71 & … & 16.99 \\
NGC3877 & 6.44 & 0.32 & … & 9.07 & 0.87 & … & 0.00 \\
NGC3893 & 1.74 & 0.46 & … & 0.57 & 0.73 & … & 4.80 \\
NGC3917 & 4 & 0.76 & … & 2.70 & 1.14 & … & 7.20 \\
NGC3949 & 1.38 & 0.46 & … & 0.76 & 0.73 & … & 6.28 \\
NGC3953 & 2.96 & 0.62 & … & 0.76 & 0.90 & … & 0.41 \\
NGC3972 & 2.94 & 0.55 & … & 2.08 & 1.03 & … & 17.63 \\
NGC3992 & 1.58 & 0.56 & … & 1.68 & 1.04 & … & 1.88 \\
NGC4010 & 4.56 & 0.46 & … & 1.90 & 0.85 & … & 38.02 \\
NGC4013 & 0.95 & 0.54 & 0.82 & 1.53 & 0.54 & 1.43 & 6.04 \\
NGC4051 & 3.53 & 0.47 & … & 1.91 & 0.81 & … & 0.98 \\
NGC4068 & 10.89 & 0.44 & … & 0.27 & 0.80 & … & 1,383.44 \\
NGC4085 & 15.12 & 0.39 & … & 2.83 & 0.58 & … & 44.82 \\
NGC4088 & 0.66 & 0.39 & … & 0.87 & 0.63 & … & 4.78 \\
NGC4100 & 1.29 & 0.57 & … & 1.90 & 0.95 & … & 3.50 \\
NGC4138 & 7.83 & 0.62 & 0.7 & 0.75 & 0.98 & 0.00 & 1.98 \\
NGC4157 & 0.57 & 0.48 & 0.67 & 0.61 & 0.69 & 0.56 & 4.08 \\
NGC4183 & 0.33 & 0.46 & … & 0.53 & 1.24 & … & 9.51 \\
NGC4214 & 1.01 & 0.54 & … & 1.19 & 1.01 & … & 42.96 \\
NGC4217 & 3.27 & 0.98 & 0.22 & 1.33 & 1.13 & 0.46 & 6.24 \\
NGC4389 & 26.19 & 0.32 & … & 0.20 & 0.31 & … & 562.73 \\
NGC4559 & 0.3 & 0.43 & … & 0.34 & 0.77 & … & 18.47 \\
NGC5005 & 0.09 & 0.5 & 0.55 & 0.07 & 0.62 & 0.69 & 1.36 \\
NGC5033 & 3.44 & 0.35 & 0.39 & 6.39 & 0.85 & 0.52 & 1.36 \\
NGC5055 & 2.54 & 0.26 & … & 6.29 & 0.62 & … & 2.12 \\
NGC5371 & 5.03 & 0.26 & … & 12.55 & 0.78 & … & 0.49 \\
NGC5585 & 6.94 & 0.21 & … & 6.35 & 0.75 & … & 149.12 \\
NGC5907 & 5.52 & 0.3 & … & 7.88 & 0.94 & … & 2.27 \\
NGC5985 & 2.38 & 0.28 & 0.65 & 3.84 & 0.00 & 1.98 & 13.73 \\
NGC6015 & 8.89 & 0.55 & … & 13.06 & 0.99 & … & 4.41 \\
NGC6195 & 2.06 & 0.36 & 0.84 & 2.95 & 0.41 & 0.81 & 1.73 \\
NGC6503 & 1.58 & 0.41 & … & 1.44 & 0.75 & … & 16.96 \\
NGC6674 & 3.42 & 1.29 & 0.92 & 3.98 & 0.70 & 1.87 & 0.68 \\
NGC6789 & NaN & NaN & NaN & 0.57 & 1.41 & … & 865.65 \\
NGC6946 & 1.7 & 0.5 & 0.53 & 1.75 & 0.64 & 0.58 & 2.52 \\
NGC7331 & 0.81 & 0.42 & 0.61 & 1.05 & 0.53 & 1.17 & 2.46 \\
NGC7793 & 0.94 & 0.55 & … & 0.75 & 0.89 & … & 12.82 \\
NGC7814 & 0.7 & 0.65 & 0.57 & 0.63 & 0.38 & 0.68 & 1.70 \\
PGC51017 & 22.6 & 0.44 & … & 2.31 & 0.80 & … & 0.00 \\
UGC00128 & 3.71 & 0.53 & … & 6.16 & 1.64 & … & 14.71 \\
UGC00191 & 6.09 & 0.58 & … & 2.45 & 1.38 & … & 35.72 \\
UGC00634 & NaN & NaN & NaN & 5.83 & 1.57 & … & 73.92 \\
UGC00731 & 0.54 & 0.55 & … & 0.08 & 3.79 & … & 26.23 \\
UGC00891 & NaN & NaN & NaN & 1.45 & 1.34 & … & 545.08 \\
UGC01230 & 1.7 & 0.52 & … & 0.80 & 1.71 & … & 7.96 \\
UGC01281 & 1.88 & 0.55 & … & 0.34 & 1.42 & … & 587.19 \\
UGC02023 & NaN & NaN & NaN & 0.06 & 0.72 & … & 1,265.61 \\
UGC02259 & 2.44 & 0.43 & … & 4.66 & 1.81 & … & 16.62 \\
UGC02455 & 10.42 & 0.34 & … & 0.90 & 0.28 & … & 843.15 \\
UGC02487 & 5.26 & 0.7 & 0.59 & 4.18 & 1.24 & 0.77 & 0.91 \\
UGC02885 & 1 & 0.42 & 1.05 & 2.14 & 0.50 & 0.98 & 1.83 \\
UGC02916 & 11.4 & 0.5 & 0.5 & 11.36 & 1.41 & 0.71 & 0.28 \\
UGC02953 & 5.64 & 0.54 & 0.57 & 5.69 & 0.74 & 0.78 & 0.92 \\
UGC03205 & 3.54 & 0.68 & 1.04 & 2.99 & 0.68 & 1.07 & 1.58 \\
UGC03546 & 1.03 & 0.59 & 0.47 & 1.14 & 0.61 & 0.60 & 1.41 \\
UGC03580 & 2.36 & 0.68 & 0.28 & 2.20 & 0.48 & 0.41 & 71.41 \\
UGC04278 & 3.19 & 0.48 & … & 0.92 & 1.00 & … & 642.15 \\
UGC04305 & 2.22 & 0.53 & … & 1.80 & 0.88 & … & 0.00 \\
UGC04325 & 6.66 & 0.49 & … & 3.72 & 1.87 & … & 0.00 \\
UGC04483 & 1.33 & 0.48 & … & 0.52 & 1.28 & … & 1,016.96 \\
UGC04499 & 1.84 & 0.47 & … & 1.54 & 1.14 & … & 94.39 \\
UGC05005 & 1.38 & 0.49 & … & 0.09 & 1.03 & … & 164.00 \\
UGC05253 & 2.5 & 0.28 & 0.75 & 5.65 & 0.53 & 0.72 & 1.37 \\
UGC05414 & 10.59 & 0.47 & … & 0.19 & 1.06 & … & 218.15 \\
UGC05716 & 2.49 & 0.45 & … & 3.81 & 1.52 & … & 148.84 \\
UGC05721 & 1.37 & 0.47 & … & 0.84 & 1.11 & … & 79.50 \\
UGC05750 & 2.08 & 0.52 & … & 0.47 & 1.57 & … & 39.99 \\
UGC05764 & 10.91 & 0.45 & … & 9.45 & 3.70 & … & 49.28 \\
UGC05829 & 0.6 & 0.51 & … & 0.07 & 1.85 & … & 145.88 \\
UGC05918 & 0.31 & 0.51 & … & 0.21 & 2.31 & … & 12.08 \\
UGC05986 & 7.1 & 0.49 & … & 1.69 & 1.15 & … & 39.53 \\
UGC05999 & NaN & NaN & NaN & 3.56 & 1.27 & … & 128.32 \\
UGC06399 & 1.75 & 0.56 & … & 0.22 & 1.43 & … & 53.24 \\
UGC06446 & 0.36 & 0.51 & … & 0.17 & 1.69 & … & 48.05 \\
UGC06614 & 0.28 & 0.47 & 0.71 & 1.20 & 0.72 & 0.59 & 7.81 \\
UGC06628 & 0.9 & 0.51 & … & 0.57 & 0.83 & … & 0.00 \\
UGC06667 & 3.46 & 0.64 & … & 2.41 & 3.77 & … & 0.00 \\
UGC06786 & 0.87 & 0.4 & 0.61 & 0.69 & 0.50 & 0.66 & 6.56 \\
UGC06787 & 18.57 & 1.1 & 0.3 & 24.78 & 0.73 & 0.57 & 2.24 \\
UGC06818 & 9.52 & 0.41 & … & 1.24 & 0.53 & … & 2,418.29 \\
UGC06917 & 1.5 & 0.51 & … & 1.09 & 1.12 & … & 39.19 \\
UGC06923 & 4.67 & 0.47 & … & 0.88 & 0.80 & … & 124.98 \\
UGC06930 & 0.54 & 0.5 & … & 0.55 & 1.25 & … & 13.66 \\
UGC06973 & 3.82 & 0.21 & 0.48 & 0.41 & 0.36 & 0.81 & 12.97 \\
UGC06983 & 0.77 & 0.54 & … & 0.77 & 1.26 & … & 28.24 \\
UGC07089 & 1.22 & 0.46 & … & 0.15 & 0.95 & … & 132.60 \\
UGC07125 & 0.76 & 0.49 & … & 1.18 & 1.08 & … & 55.91 \\
UGC07151 & 3.73 & 0.72 & … & 1.30 & 1.15 & … & 32.08 \\
UGC07232 & NaN & NaN & NaN & 0.76 & 0.82 & … & 1,563.21 \\
UGC07261 & 0.87 & 0.52 & … & 1.46 & 1.22 & … & 40.71 \\
UGC07323 & 2.63 & 0.49 & … & 0.26 & 0.97 & … & 86.65 \\
UGC07399 & 1.99 & 0.54 & … & 1.00 & 1.54 & … & 70.39 \\
UGC07524 & 0.97 & 0.5 & … & 1.48 & 1.59 & … & 34.22 \\
UGC07559 & 1.97 & 0.43 & … & 0.21 & 0.97 & … & 1,061.77 \\
UGC07577 & 3.11 & 0.37 & … & 0.06 & 0.67 & … & 4,668.25 \\
UGC07603 & 2.36 & 0.46 & … & 0.52 & 1.07 & … & 288.77 \\
UGC07608 & 1.94 & 0.51 & … & 0.30 & 2.02 & … & 428.16 \\
UGC07690 & 3.5 & 0.67 & … & 0.40 & 1.05 & … & 29.31 \\
UGC07866 & 0.3 & 0.46 & … & 0.07 & 1.26 & … & 330.26 \\
UGC08286 & 3.02 & 0.62 & … & 2.66 & 1.66 & … & 36.50 \\
UGC08490 & 0.29 & 0.48 & … & 0.15 & 1.26 & … & 40.48 \\
UGC08550 & 1.53 & 0.6 & … & 0.70 & 1.35 & … & 171.01 \\
UGC08699 & 0.72 & 0.74 & 0.59 & 0.95 & 0.55 & 0.76 & 1.94 \\
UGC08837 & 10.79 & 0.38 & … & 0.69 & 0.78 & … & 1,422.40 \\
UGC09037 & 2.13 & 0.25 & … & 1.55 & 0.58 & … & 26.53 \\
UGC09133 & 7.17 & 0.75 & 0.53 & 7.50 & 0.77 & 0.72 & 0.71 \\
UGC09992 & NaN & NaN & NaN & 0.04 & 1.30 & … & 0.10 \\
UGC10310 & 1.63 & 0.51 & … & 0.12 & 1.54 & … & 19.21 \\
UGC11455 & 5.47 & 0.61 & … & 3.98 & 0.76 & … & 2.41 \\
UGC11557 & 2.06 & 0.47 & … & 0.91 & 0.60 & … & 73.80 \\
UGC11820 & 2.63 & 0.58 & … & 1.44 & 1.24 & … & 145.34 \\
UGC11914 & 0.87 & 0.34 & 0.91 & 1.61 & 0.04 & 0.89 & 2.46 \\
UGC12506 & 0.26 & 0.43 & … & 1.18 & 1.35 & … & 1.20 \\
UGC12632 & 0.45 & 0.48 & … & 0.21 & 1.86 & … & 31.86 \\
UGC12732 & 0.26 & 0.5 & … & 0.18 & 1.51 & … & 82.38 \\
UGCA281 & 1.46 & 0.5 & … & 0.28 & 1.23 & … & 339.55 \\
UGCA442 & 7.38 & 0.61 & … & 0.86 & 2.28 & … & 261.55 \\
UGCA444 & 0.23 & 0.51 & … & 0.10 & 3.78 & … & 425.75 \\
 \end{longtable}
\end{center}


\begin{table*}[!ht]
 \caption{ \textbf{Best-fit Parameters for the most reliable SPARC Galaxy Data}\\
 \emph{Column 2} Galaxies for which the dark matter  model fails to find a  fit are indicated by NaN.  \\
\emph{Columns 3 \& 4} Our model results with   MW baseline from  McGaugh-Jiao \cite{McGaugh_2019,jiao2023detection} indicated by  \textbf{MJ}.\\
 \emph{Columns 5 \& 6} Our model results with   MW baseline from
 from Xue-Sofue-Jiao \cite{Sofue,Xue,jiao2023detection} indicated by 
 \textbf{XSJ}. \\
 \emph{Columns 7 \& 8}  The reported distances   and the  total luminosity $L$ in units of $10^9 M_{\odot}$ by  
the   half-light radius  $R$  in units of  $kpc$,  as reported in   the SPARC database \citep{2016Lelli}.
 \label{TSet}} 
\begin{tabular}{|l|c|rr|rr|r|r|}
\toprule
\hline \multicolumn{1}{|c|}{  } 
& \multicolumn{1}{c|}{\textbf{Dark Matter}} 
& \multicolumn{2}{c|}{  Our Model   }  
& \multicolumn{2}{c|}{  Our Model    }
& \multicolumn{1}{c|}{  }
& \multicolumn{1}{c|}{}\\  
 \multicolumn{1}{|c|}{ } 
& \multicolumn{1}{c|}{\textbf{NFW}} 
& \multicolumn{2}{c|}{     \textbf{MJ } }  
& \multicolumn{2}{c|}{      \textbf{XSJ } }
& \multicolumn{1}{c|}{  Distance}
& \multicolumn{1}{c|}{ \textbf{ $L/R$}}\\     
 Name &$\chi^2_{r}$  & $\chi^2_{r}$& $\alpha$& $\chi^2_{r}$ & $\alpha$         &(Mpc)    &     \\  
\hline  
CamB & 7.92 & 0.81 & 2,505.76 & 0.23 & 24,989.21 & 3.36 & 0.06 \\ \hline 
D564-8 & 6.70 & 0.03 & 1,177.28 & 0.11 & 4,130.26 & 8.79 & 0.05 \\ \hline 
D631-7 & 12.21 & 1.22 & 501.02 & 0.30 & 2,725.54 & 7.72 & 0.16 \\ \hline 
DDO154 & 16.57 & 6.15 & 500.74 & 10.58 & 1,567.39 & 4.04 & 0.08 \\ \hline 
DDO168 & 23.93 & 4.55 & 330.48 & 4.36 & 1,290.43 & 4.25 & 0.15 \\ \hline 
ESO444-G084 & 4.71 & 1.19 & 156.01 & 0.40 & 465.01 & 4.83 & 0.09 \\ \hline 
IC2574 & 36.30 & 1.69 & 233.95 & 2.27 & 885.47 & 3.91 & 0.32 \\ \hline 
NGC0024 & 0.66 & 0.94 & 2.96 & 0.73 & 9.88 & 7.30 & 1.93 \\ \hline 
NGC0055 & 4.35 & 1.51 & 35.98 & 2.86 & 106.90 & 2.11 & 1.26 \\ \hline 
NGC0247 & 2.03 & 2.36 & 1.78 & 2.18 & 8.03 & 3.70 & 1.25 \\ \hline 
NGC0300 & 0.93 & 0.44 & 33.21 & 0.42 & 85.28 & 2.08 & 1.65 \\ \hline 
NGC2403 & 9.09 & 14.13 & 9.06 & 10.73 & 19.32 & 3.16 & 4.65 \\ \hline 
NGC2683 & 2.66 & 0.96 & 0.85 & 1.01 & 1.78 & 9.81 & 24.08 \\ \hline 
NGC2841 & 1.47 & 1.19 & 2.20 & 1.36 & 1.40 & 14.10 & 34.14 \\ \hline 
NGC2915 & 1.05 & 0.59 & 917.03 & 0.63 & 841.91 & 4.06 & 1.21 \\ \hline 
NGC2976 & 1.45 & 0.47 & 6.77 & 0.46 & 40.19 & 3.58 & 2.57 \\ \hline 
NGC3109 & 15.31 & 0.35 & 164.55 & 0.28 & 661.98 & 1.33 & 0.12 \\ \hline 
NGC3198 & 1.46 & 1.31 & 3.88 & 1.72 & 8.51 & 13.80 & 6.55 \\ \hline 
NGC3741 & 1.74 & 0.60 & 2,271.94 & 0.65 & 6,628.19 & 3.21 & 0.09 \\ \hline 
NGC4214 & 1.01 & 1.60 & 15.38 & 1.19 & 42.96 & 2.87 & 1.63 \\ \hline 
NGC5055 & 2.54 & 7.70 & 1.28 & 6.29 & 2.12 & 9.90 & 36.58 \\ \hline 
NGC6503 & 1.58 & 1.72 & 9.31 & 1.44 & 16.96 & 6.26 & 7.93 \\ \hline 
NGC6789 & NaN & 0.08 & 2,993.97 & 0.57 & 865.65 & 3.52 & 0.19 \\ \hline 
NGC6946 & 1.70 & 2.52 & 1.74 & 1.75 & 2.52 & 5.52 & 15.76 \\ \hline 
NGC7331 & 0.81 & 1.04 & 1.70 & 1.05 & 2.46 & 14.70 & 62.81 \\ \hline 
NGC7793 & 0.94 & 0.71 & 4.21 & 0.75 & 12.82 & 3.61 & 3.22 \\ \hline 
UGC04483 & 1.33 & 0.45 & 214.19 & 0.52 & 1,016.96 & 3.34 & 0.05 \\ \hline 
UGC07232 & NaN & 0.85 & 488.75 & 0.76 & 1,563.21 & 2.83 & 0.25 \\ \hline 
UGC07524 & 0.97 & 1.28 & 14.09 & 1.48 & 34.22 & 4.74 & 0.67 \\ \hline 
UGC07559 & 1.97 & 0.17 & 203.80 & 0.21 & 1,061.77 & 4.97 & 0.11 \\ \hline 
UGC07577 & 3.11 & 0.06 & 737.74 & 0.06 & 4,668.25 & 2.59 & 0.06 \\ \hline 
UGC07866 & 0.30 & 0.08 & 55.68 & 0.07 & 330.26 & 4.57 & 0.13 \\ \hline 
UGC08490 & 0.29 & 0.26 & 17.67 & 0.15 & 40.48 & 4.65 & 0.89 \\ \hline 
UGC08837 & 10.79 & 0.61 & 309.13 & 0.69 & 1,422.40 & 7.21 & 0.22 \\ \hline 
UGCA442 & 7.38 & 0.72 & 74.81 & 0.86 & 261.55 & 4.35 & 0.08 \\ \hline 
UGCA444 & 0.23 & 0.12 & 98.80 & 0.10 & 425.75 & 0.98 & 0.03 \\ \hline 
\end{tabular}
\end{table*}
  
\clearpage
 
\section[\appendixname~\thesection]{  Individual Galaxy Comparisons}
\label{results:MtoL}

In this appendix we highlight the Universal Rotation Curve Spectrum groupings \cite{PSS}   the three classes of rotation curves; declining, flat and ascending.  
 \subsection[\appendixname~\thesubsection]{   Declining Rotation Curves}

\begin{figure}[!ht] 
\begin{adjustwidth}{-\extralength}{0cm}
\centering 
\subfloat[\centering]{\includegraphics[width=6.0cm]{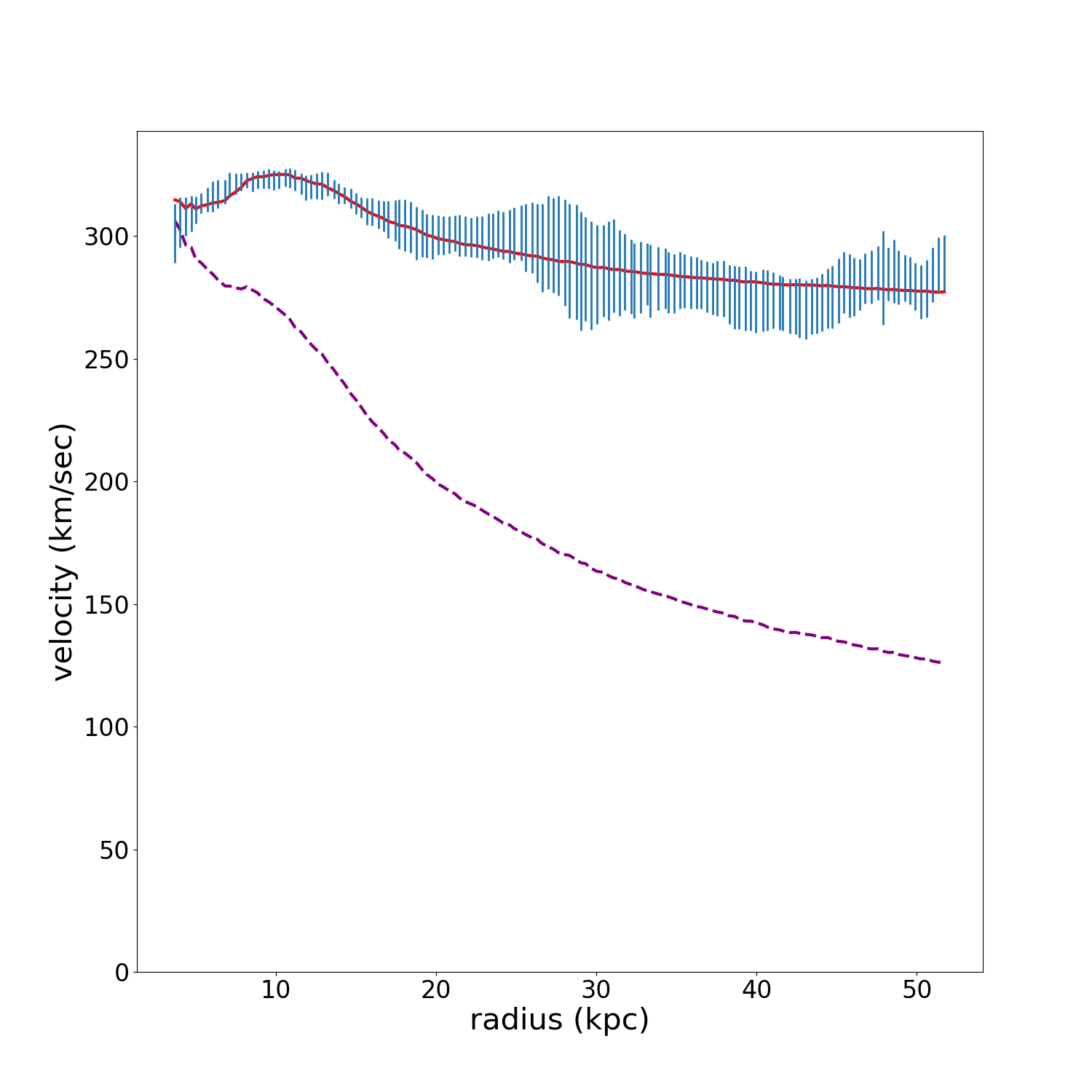}}
\subfloat[\centering]{\includegraphics[width=6.0cm]{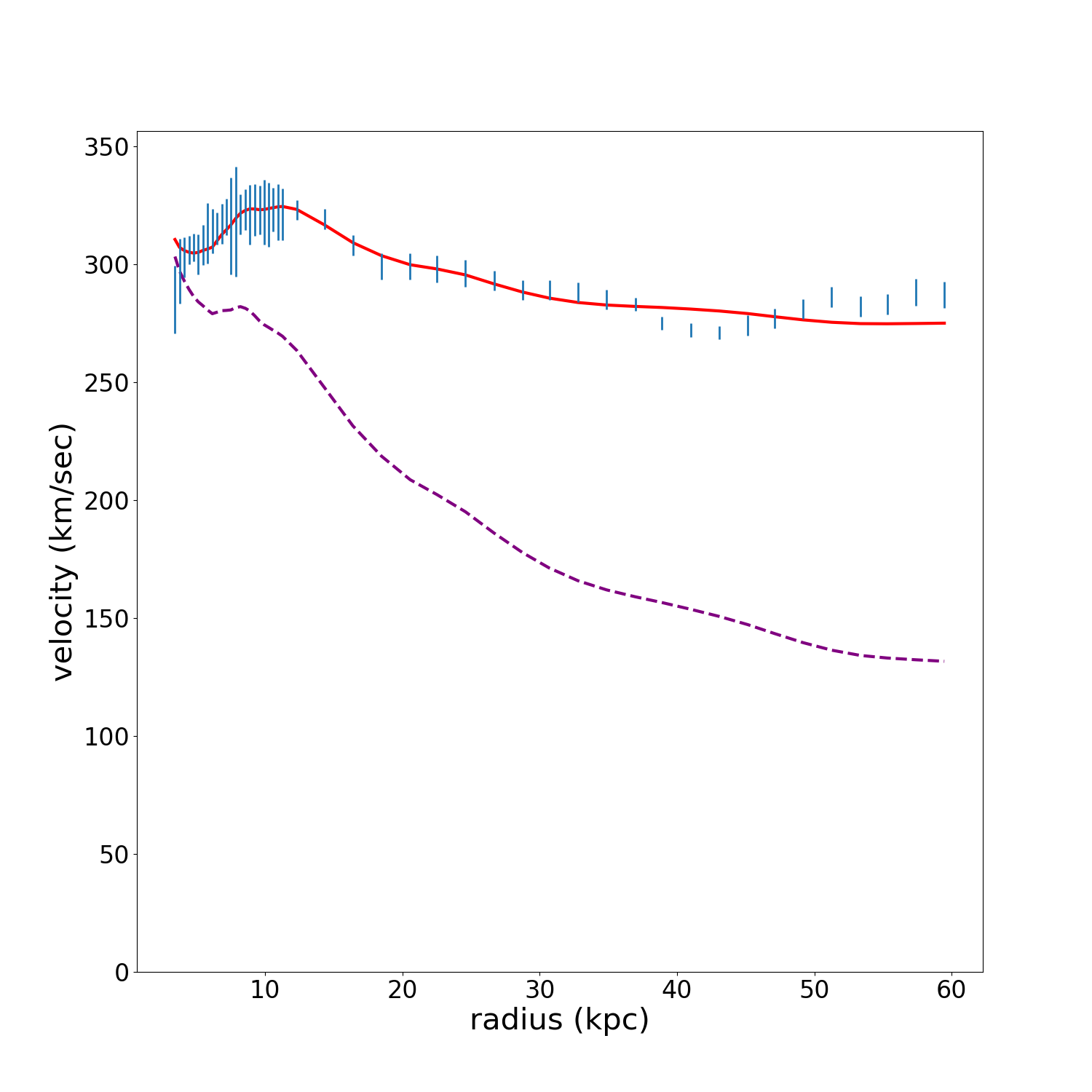}}\\
\subfloat[\centering]{\includegraphics[width=6.0cm]{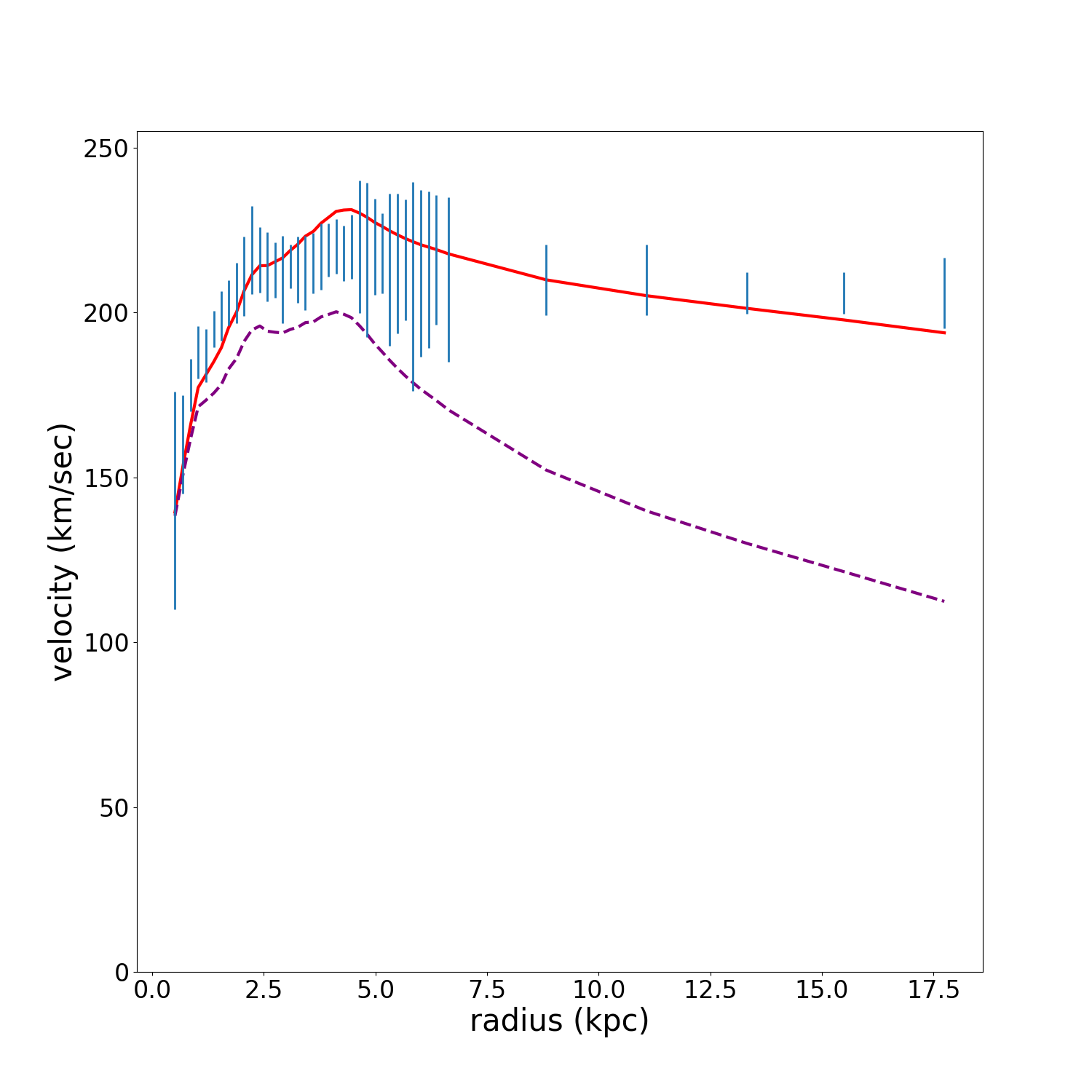}}
\subfloat[\centering]{\includegraphics[width=6.0cm]{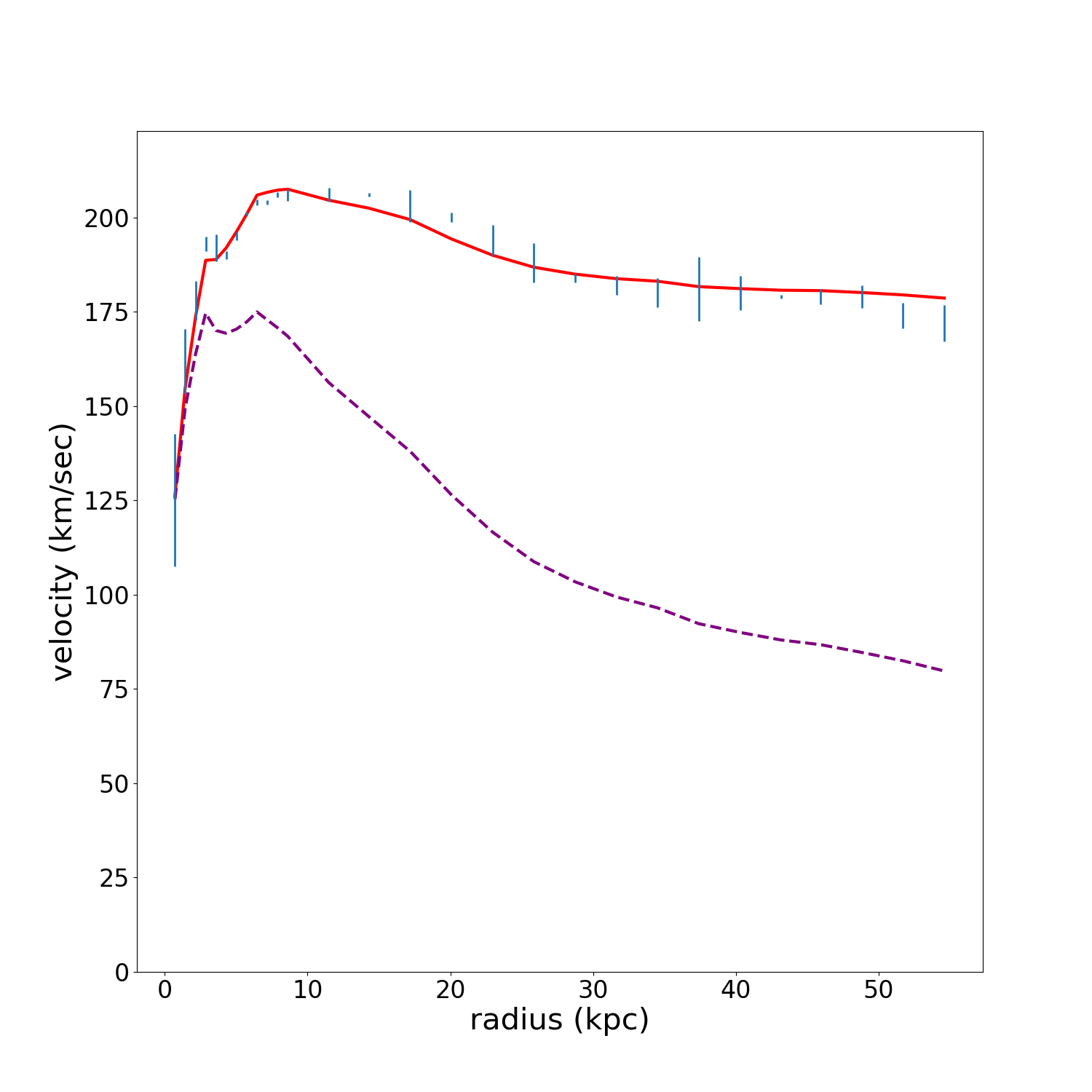 }}\\
\subfloat[\centering]{\includegraphics[width=6.0cm]{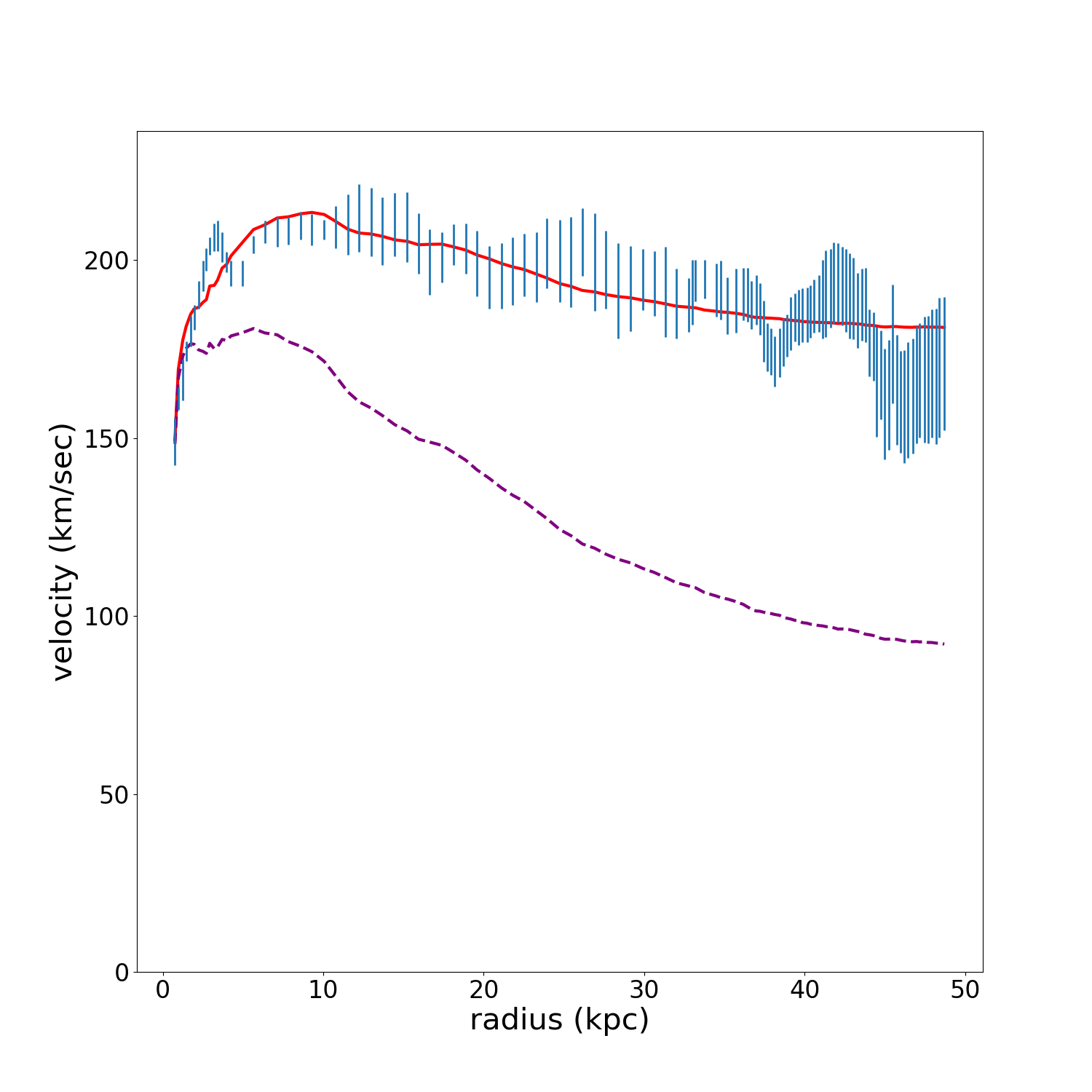 }}
\end{adjustwidth} 
\caption{ Declining Rotation Curves of high-surface brightness galaxies. Rotation curve data (blue points with error bars) and the input baryonic mass model (purple dashed line).   This   model's fits are  shown by the red solid line, with respect to the Xue-Sofue-Jiao Milky Way.  
(\textbf{a}) NGC 2841 \cite{Blok1},
(\textbf{b}) NGC 2841 \cite{2016Lelli},
(\textbf{c}) NGC 3521 \cite{2016Lelli},
(\textbf{d}) NGC 5055 \cite{2016Lelli},
(\textbf{e}) NGC 5055 \cite{Blok1}.
\label{N2841a} }
\end{figure}

 As can be seen in Figure~\ref{N2841a},  our model provides excellent      fits to the well studied, high-surface brightness,   spiral galaxies NGC 2841, 
 NGC 5055,  and NGC 3521.  
 \citet{TWiegert} has noted that these are  galaxies with declining rotation curves that are not well explained by  the usual dark matter morphology scenarios.
  The   under-constrained nature of luminous mass modeling   \citep{Conroy} is highlighted for NGC 5055, with a    luminous mass model   from  \citet{Batt,2016Lelli} with no   central stellar bulge, and one from   \citet{Blok1} which  does.

\subsection[\appendixname~\thesubsection]{ Truly Flat Rotation Curves}
\begin{figure}[!ht] 
\begin{adjustwidth}{-\extralength}{0cm}
\centering 
\subfloat[\centering]{\includegraphics[width=6.0cm]{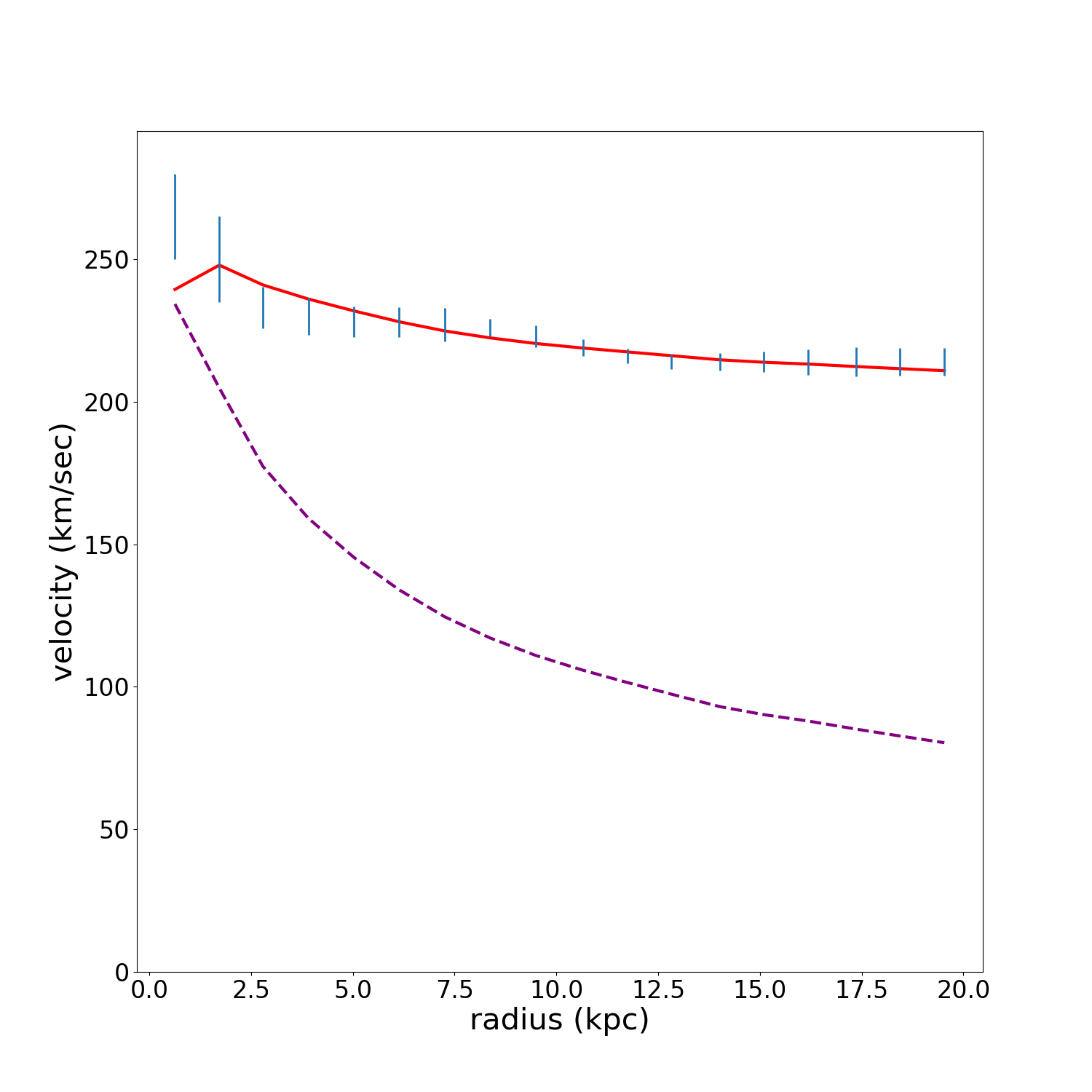}}
\subfloat[\centering]{\includegraphics[width=6.0cm]{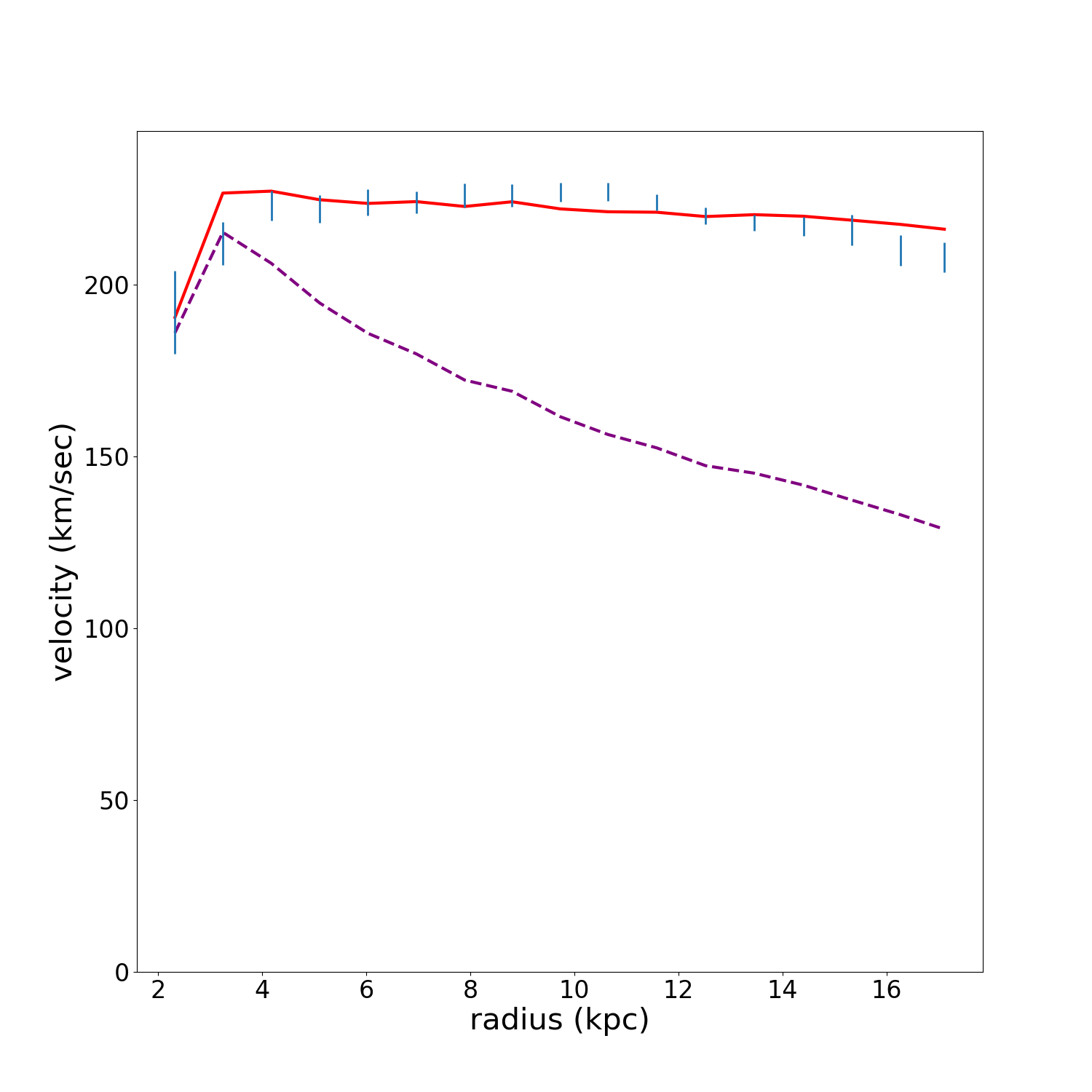}}\\
\subfloat[\centering]{\includegraphics[width=6.0cm]{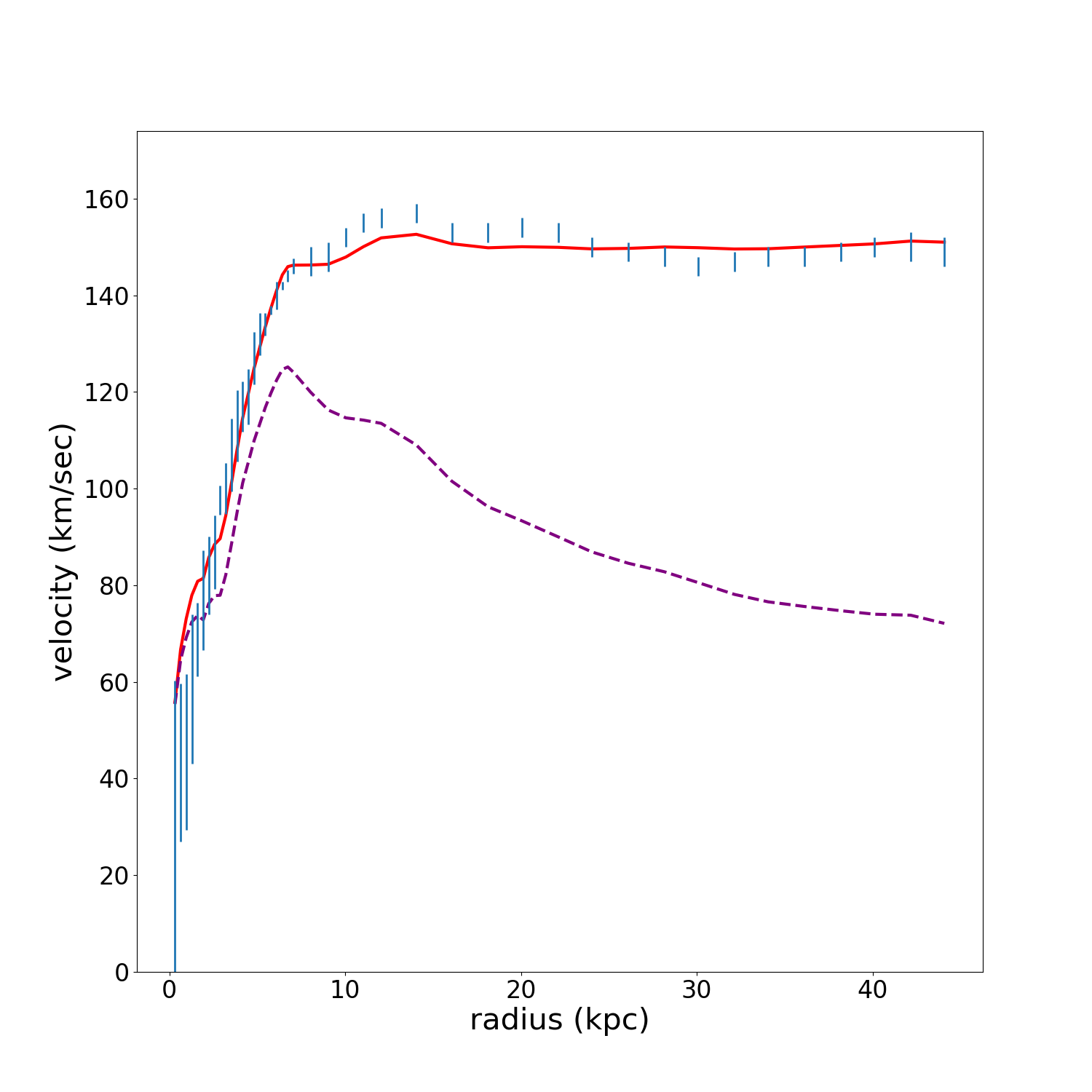}}
\subfloat[\centering]{\includegraphics[width=6.0cm]{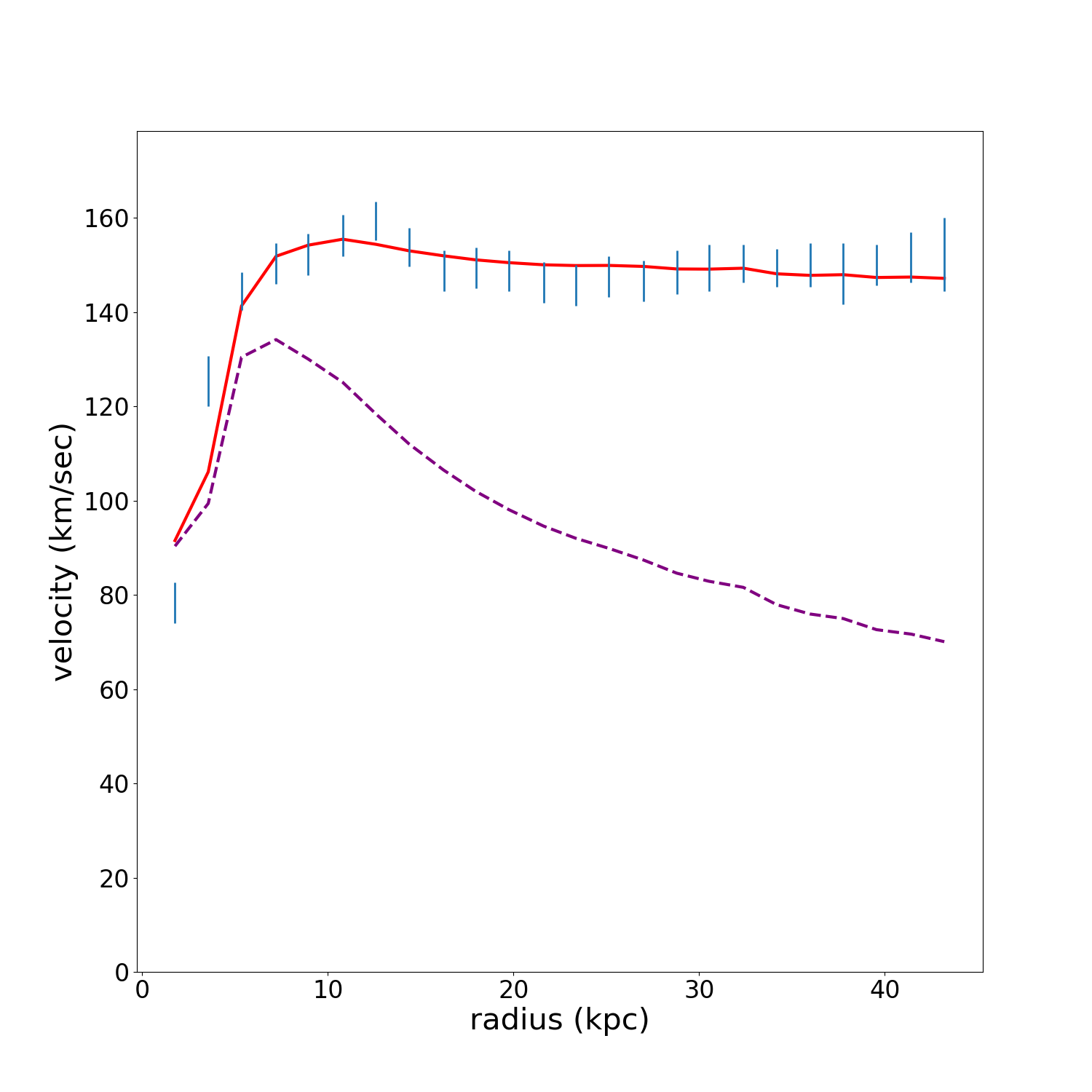}}\\
\end{adjustwidth} 
\caption{ Flat Rotation Curves.   Lines are as in Fig.~\ref{N2841a} 
(\textbf{a}) NGC 7814 \cite{Frat1,2016Lelli},
(\textbf{b}) NGC 891 \cite{Frat1,2016Lelli},
(\textbf{c}) NGC 3198  \cite{2016Lelli},
(\textbf{d})  NGC 3198  \cite{Gent}.
\label{fitCompare7814} }
\end{figure}

 s can be seen in  
  Fig.~\ref{fitCompare7814}, our model provides excellent fits to  galaxies with truly flat rotation curves.  We include NGC 3198, whose rotation curve is truly flat in comparison to NGC 891 and NGC 7814, whose rotation curves inflect about flat.  NGC 3198 is a barred spiral in Ursa Major, which has  no bulge component.
 The spiral galaxies NGC 7814  and NGC 891
  present an interesting test case to   dark matter   models, because both galaxies are presented edge-on on the sky and   have essentially identical RCs, but   extreme opposite    luminous mass morphologies. 
  NGC 7814 is  bulge dominated   and NGC 891 is an almost entirely  disk galaxy. In
 \cite{Frat1} the question is raised,  ``why are these RCs so identical if their dark matter halos are necessarily different to accommodate the differences in the luminous mass?'' In the paradigm of this paper's model     the two RCs are  very similar in magnitude,  but   differ  in their inflections  at large radii. In fact, it is such a subtle difference that it particularly highlights the power of this framing of the rotation curve problem.

  \subsection[\appendixname~\thesubsection]{  Ascending Rotation Curves}
\begin{figure}[!ht] 
\begin{adjustwidth}{-\extralength}{0cm}
\centering 
\subfloat[\centering]{\includegraphics[width=6.0cm]{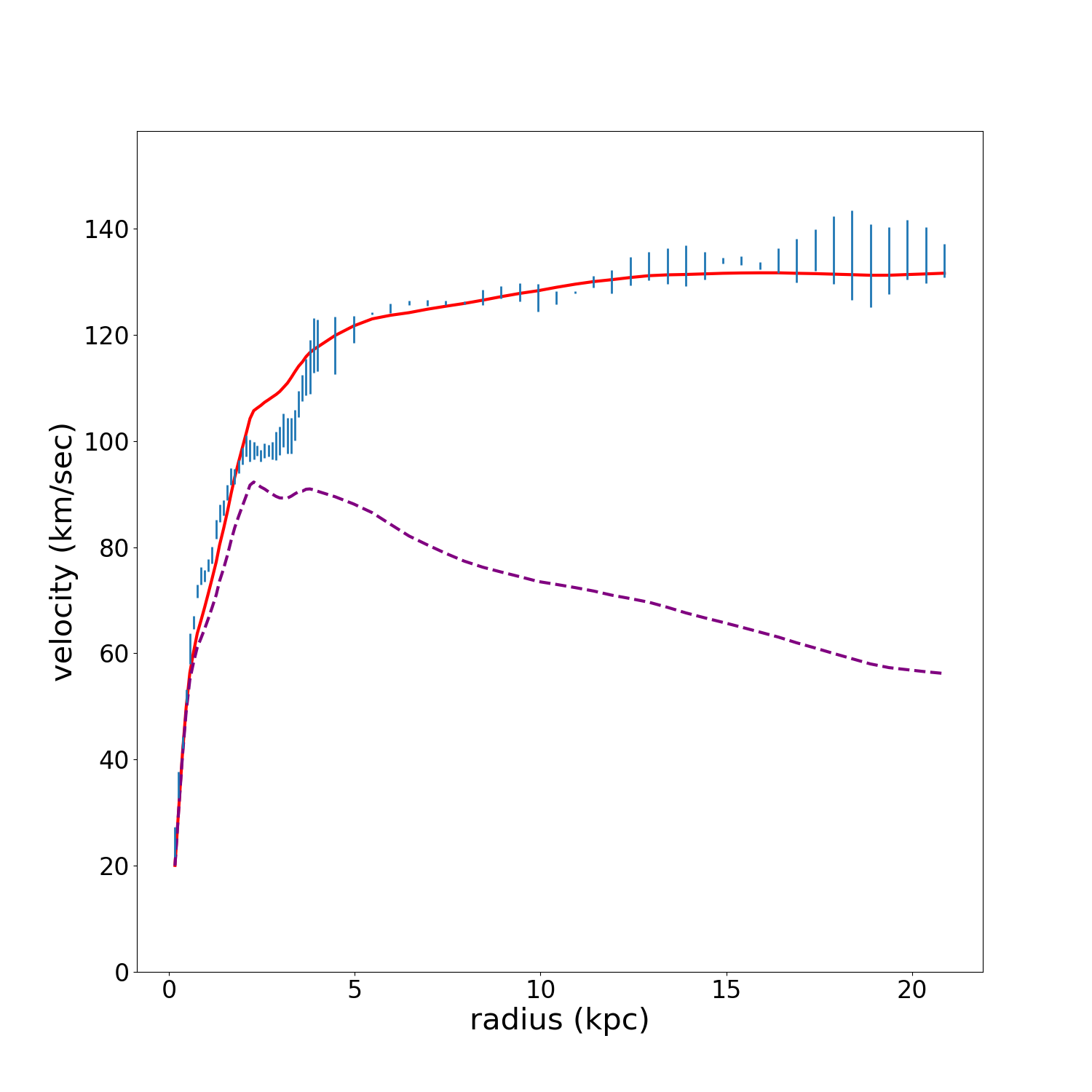}}
\subfloat[\centering]{\includegraphics[width=6.0cm]{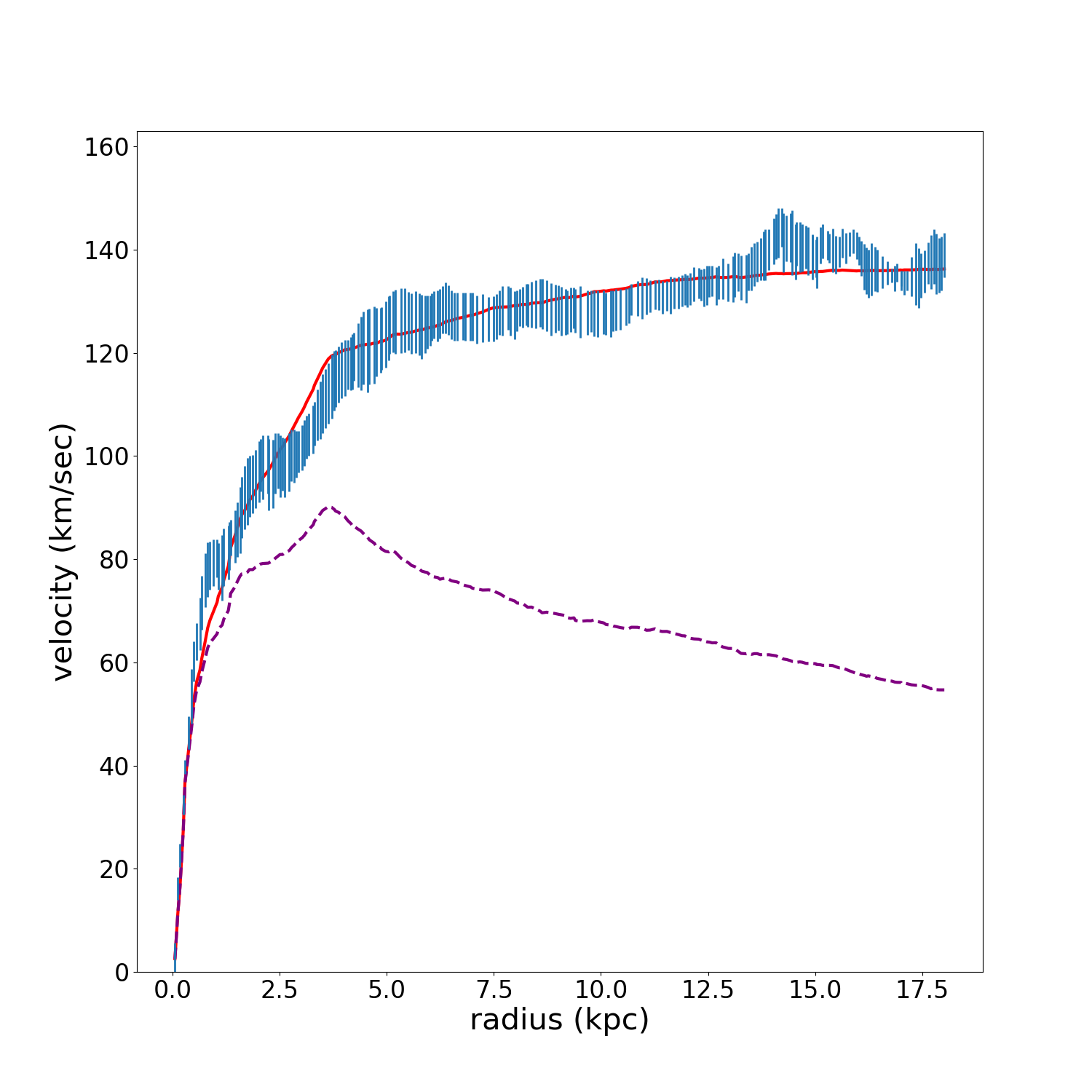}}\\
\subfloat[\centering]{\includegraphics[width=6.0cm]{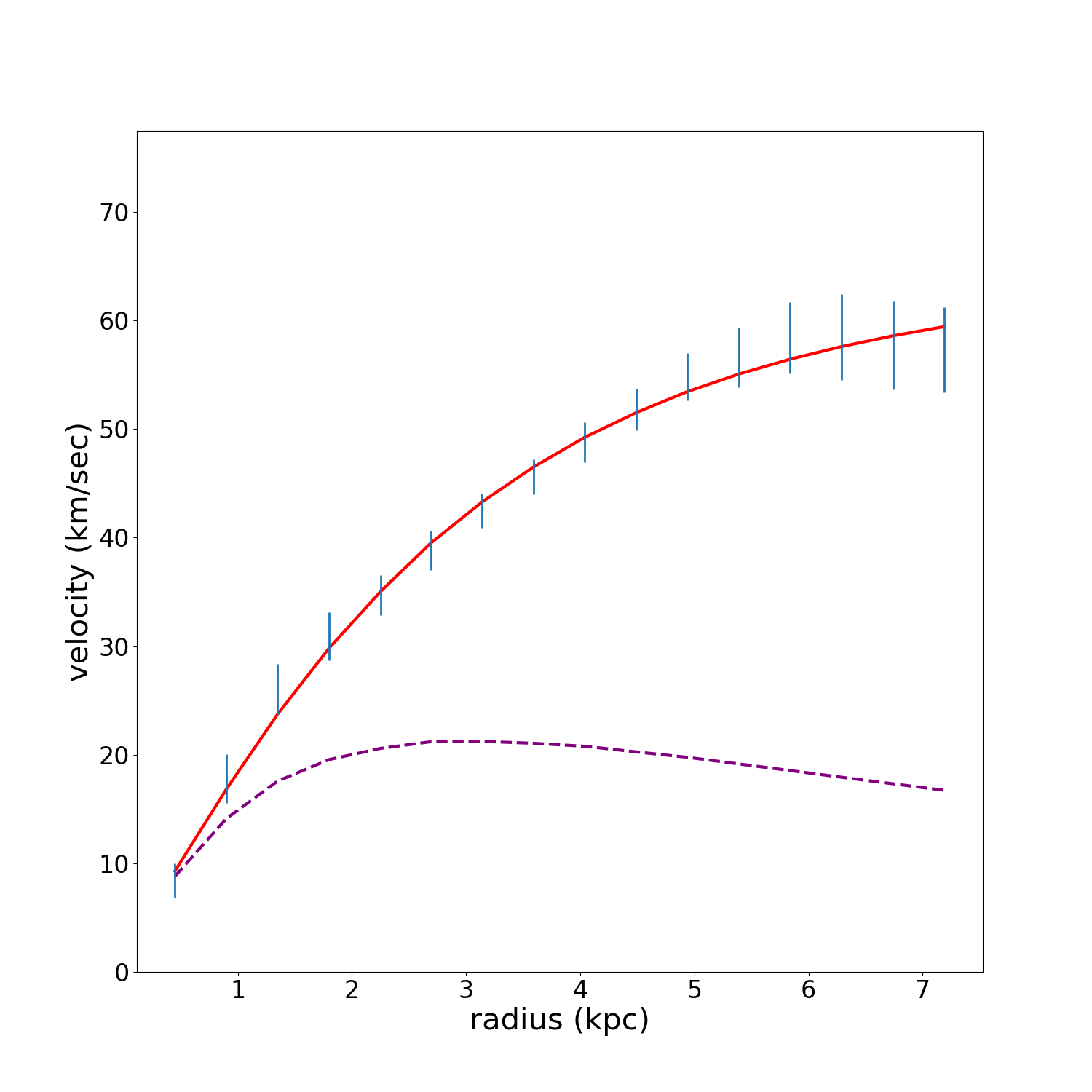}}
\subfloat[\centering]{\includegraphics[width=6.0cm]{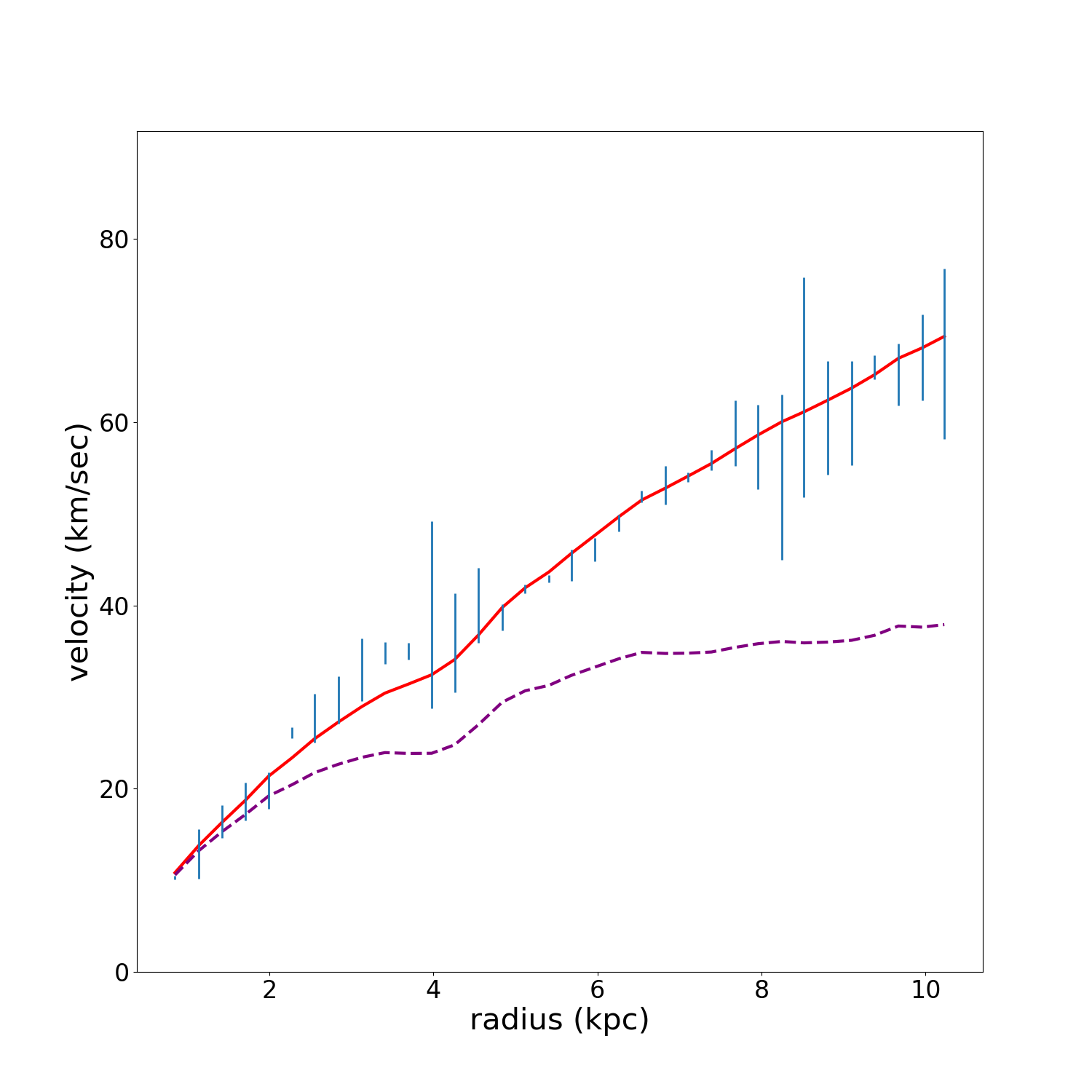}}\\
\end{adjustwidth} 
\caption{ Ascending Rotation Curves.   Lines are as in Fig.~\ref{N2841a} 
(\textbf{a}) NGC 2403 \cite{2016Lelli},
(\textbf{b}) NGC 2403 \cite{Blok1},
(\textbf{c})  D 631-7   \cite{2016Lelli},
(\textbf{d}) IC 2574 \cite{2016Lelli}.
\label{n2403} }
\end{figure}

 As can be seen in  Fig.~\ref{n2403}, our model provides excellent fits to galaxies with ascending rotation curves - including examples of both gas dominated dwarf galaxies and intermediate spirals.
Dwarf galaxies have been an important test case for dark matter models, giving rise to the cusp-core problem \cite{10.1046/j.1365-8711.2001.04456.x}. We present fit results here for two such galaxies. 
D 631-7, a dwarf irregular 
also known as UGC 4115, and   IC 2574,  a gas-dominated dwarf spiral galaxy,  with no central stellar bulge. 
IC 2574   is   problematic for dark matter model fits, as the   model   overestimates  the inner RC out to 10 kpc \citep{2017MNRAS.471.1841N}.

\clearpage

\begin{adjustwidth}{-\extralength}{0cm}

\reftitle{References}




\bibliography{LCM}

\begin{thebibliography}{999}

\bibitem[{Rubin} and {Ford}(1970)]{1970ApJ...159..379R}
{Rubin}, V.C.; {Ford}, Jr., W.K.
\newblock Rotation of the Andromeda Nebula from a Spectroscopic Survey of Emission Regions.
\newblock {\em The Astrophysical Journal} {\bf 1970}, {\em 159},~379.
\newblock {\url{https://doi.org/10.1086/150317}}.

\bibitem[Rubin et~al.(1978)Rubin, Thonnard, and Ford]{1978Rubin}
Rubin, V.C.; Thonnard, N.; Ford, Jr., W.K.
\newblock {Extended rotation curves of high-luminosity spiral galaxies. IV - Systematic dynamical properties, SA through SC}.
\newblock {\em The Astrophysical Journal} {\bf 1978}, {\em 225},~L107--L111.
\newblock {\url{https://doi.org/10.1086/182804}}.

\bibitem[Bosma(1981)]{Bosma}
Bosma, A.
\newblock The distribution and kinematics of neutral hydrogen in spiral galaxies of various morphological types.
\newblock {\em The Astronomical Journal} {\bf 1981}, {\em 86},~1791.

\bibitem[{van Albada} et~al.(1985){van Albada}, {Bahcall}, {Begeman}, and {Sancisi}]{1985ApJAlbada}
{van Albada}, T.S.; {Bahcall}, J.N.; {Begeman}, K.; {Sancisi}, R.
\newblock {Distribution of dark matter in the spiral galaxy NGC 3198}.
\newblock {\em The Astrophysical Journal} {\bf 1985}, {\em 295},~305--313.
\newblock {\url{https://doi.org/10.1086/163375}}.

\bibitem[Cebri\'an(2022)]{Cebrian:2022brv}
Cebri\'an, S.
\newblock {Review on dark matter searches}.
\newblock In Proceedings of the {10th Symposium on Large TPCs for Low-Energy Rare Event Detection},  5 2022,  \href{http://arxiv.org/abs/2205.06833}{{\normalfont [arXiv:physics.ins-det/2205.06833]}}.

\bibitem[Misiaszek and Rossi(2024)]{Misiaszek_2024}
Misiaszek, M.; Rossi, N.
\newblock Direct Detection of Dark Matter: A Critical Review.
\newblock {\em Symmetry} {\bf 2024}, {\em 16},~201.
\newblock {\url{https://doi.org/10.3390/sym16020201}}.

\bibitem[Akerib et~al.(2022)Akerib, Cushman, Dahl, Ebadi, Fan, Gaitskell, Galbiati, Giovanetti, Gelmini, Grandi, Haselschwardt, Jackson, Lang, Loer, Loomba, Marshall, Mills, OHare, Savarese, Schueler, Szydagis, Takhistov, Tait, Tsai, Vahsen, Walsworth, and Westerdale]{neutrino_fog_SNOWMASS}
Akerib, D.S.; Cushman, P.B.; Dahl, C.E.; Ebadi, R.; Fan, A.; Gaitskell, R.J.; Galbiati, C.; Giovanetti, G.K.; Gelmini, G.B.; Grandi, L.;  et~al.
\newblock Snowmass2021 Cosmic Frontier Dark Matter Direct Detection to the Neutrino Fog,  2022,  \href{http://arxiv.org/abs/2203.08084}{{\normalfont [arXiv:hep-ex/2203.08084]}}.

\bibitem[{Ostriker} and {Peebles}(1973)]{1973ApJ...186..467O}
{Ostriker}, J.P.; {Peebles}, P.J.E.
\newblock {A Numerical Study of the Stability of Flattened Galaxies: or, can Cold Galaxies Survive?}
\newblock {\em The Astrophysical Journal} {\bf 1973}, {\em 186},~467--480.
\newblock {\url{https://doi.org/10.1086/152513}}.

\bibitem[Rubin et~al.(1980)Rubin, Ford, and Thonnard]{Rub}
Rubin, V.; Ford, W.; Thonnard, N.
\newblock Rotational properties of 21 SC galaxies with a large range of luminosities and radii.
\newblock {\em The Astrophysical Journal} {\bf 1980}, {\em 238},~471.

\bibitem[Persic and Salucci(1997)]{PSS}
Persic, M.; Salucci, P.
\newblock Dark and visible matter in Galaxies.
\newblock In Proceedings of the Dark and visible matter in Galaxies. ASP Conference Series 117, 1,  1997.

\bibitem[Sofue and Rubin(2001)]{SofRub}
Sofue, Y.; Rubin, V.
\newblock Rotation Curves of Spiral Galaxies.
\newblock {\em Annual Review of Astronomy and Astrophysics} {\bf 2001}, {\em 39},~137.

\bibitem[McGaugh et~al.(2000)McGaugh, Schombert, Bothun, and de~Blok]{McGaugh_2000}
McGaugh, S.S.; Schombert, J.M.; Bothun, G.D.; de~Blok, W.J.G.
\newblock The Baryonic Tully-Fisher Relation.
\newblock {\em The Astrophysical Journal} {\bf 2000}, {\em 533},~L99–L102.
\newblock {\url{https://doi.org/10.1086/312628}}.

\bibitem[McGaugh(1999)]{1999McGaugh}
McGaugh, S.
\newblock Galaxy Dynamics.
\newblock {\em Conference Proceedings} {\bf 1999}, {\em 182}.

\bibitem[{McGaugh}(2004)]{2004ApJ...609..652M}
{McGaugh}, S.S.
\newblock {The Mass Discrepancy-Acceleration Relation: Disk Mass and the Dark Matter Distribution}.
\newblock {\em The Astrophysical Journal} {\bf 2004}, {\em 609},~652--666,  \href{http://arxiv.org/abs/astro-ph/0403610}{{\normalfont [arXiv:astro-ph/astro-ph/0403610]}}.
\newblock {\url{https://doi.org/10.1086/421338}}.

\bibitem[{Tully} and {Fisher}(1977)]{1977A&A....54..661T}
{Tully}, R.B.; {Fisher}, J.R.
\newblock {A new method of determining distances to galaxies.}
\newblock {\em Astronomy and Astrophysics} {\bf 1977}, {\em 54},~661--673.

\bibitem[Ziegler et~al.(2001)Ziegler, Böhm, Fricke, Jäger, Nicklas, Bender, Drory, Gabasch, Saglia, Seitz, Heidt, Mehlert, Möllenhoff, Noll, and Sutorius]{Ziegler_2002}
Ziegler, B.L.; Böhm, A.; Fricke, K.J.; Jäger, K.; Nicklas, H.; Bender, R.; Drory, N.; Gabasch, A.; Saglia, R.P.; Seitz, S.;  et~al.
\newblock The Evolution of the Tully-Fisher Relation of Spiral Galaxies*.
\newblock {\em The Astrophysical Journal} {\bf 2001}, {\em 564},~L69.
\newblock {\url{https://doi.org/10.1086/338962}}.

\bibitem[{Persic} et~al.(1996){Persic}, {Salucci}, and {Stel}]{salucci}
{Persic}, M.; {Salucci}, P.; {Stel}, F.
\newblock {The universal rotation curve of spiral galaxies - I. The dark matter connection}.
\newblock {\em Monthly Notices of the Royal Astronomical Society} {\bf 1996}, {\em 281},~27--47,  \href{http://arxiv.org/abs/astro-ph/9506004}{{\normalfont [astro-ph/9506004]}}.

\bibitem[Salucci and Burkert(2000)]{SalucciApJ}
Salucci, P.; Burkert, A.
\newblock Dark Matter Scaling Relations.
\newblock {\em The Astrophysical Journal} {\bf 2000}, {\em 537},~L9--L12.

\bibitem[Maschberger et~al.(2014)Maschberger, Bonnell, Clarke, and Moraux]{10.1093/mnras/stt2403}
Maschberger, T.; Bonnell, I.A.; Clarke, C.J.; Moraux, E.
\newblock {The relation between accretion rates and the initial mass function in hydrodynamical simulations of star formation}.
\newblock {\em Monthly Notices of the Royal Astronomical Society} {\bf 2014}, {\em 439},~234--246,  \href{http://arxiv.org/abs/https://academic.oup.com/mnras/article-pdf/439/1/234/5571048/stt2403.pdf}{{\normalfont [https://academic.oup.com/mnras/article-pdf/439/1/234/5571048/stt2403.pdf]}}.
\newblock {\url{https://doi.org/10.1093/mnras/stt2403}}.

\bibitem[Iocco et~al.(2015)Iocco, Pato, and Bertone]{Iocco_2015}
Iocco, F.; Pato, M.; Bertone, G.
\newblock Evidence for dark matter in the inner Milky Way.
\newblock {\em Nature Physics} {\bf 2015}, {\em 11},~245–248.
\newblock {\url{https://doi.org/10.1038/nphys3237}}.

\bibitem[Li et~al.(2020)Li, Lelli, McGaugh, and Schombert]{Li_2020}
Li, P.; Lelli, F.; McGaugh, S.; Schombert, J.
\newblock A Comprehensive Catalog of Dark Matter Halo Models for SPARC Galaxies.
\newblock {\em The Astrophysical Journal Supplement Series} {\bf 2020}, {\em 247},~31.
\newblock {\url{https://doi.org/10.3847/1538-4365/ab700e}}.

\bibitem[Spinrad and Taylor(1971)]{1971ApJS...22..445S}
Spinrad, H.; Taylor, B.J.
\newblock {The Stellar Content of the Nuclei of Nearby Galaxies. I. M31, M32, and M81}.
\newblock {\em The Astrophysical Journal} {\bf 1971}, {\em 22},~445.
\newblock {\url{https://doi.org/10.1086/190232}}.

\bibitem[Conroy et~al.(2009)Conroy, Gunn, and White]{Conroy}
Conroy, C.; Gunn, J.; White, M.
\newblock The Propagation of Uncertainties in stellar population synthesis modeling I: The relevance of uncertain aspects of stellar evolution and the IMF to the derived physical properties of galaxies.
\newblock {\em The Astrophysical Journal} {\bf 2009}, {\em 699},~486.

\bibitem[Munshi et~al.(2013)Munshi, Governato, Brooks, Christensen, Shen, Loebman, Moster, Quinn, and Wadsley]{Munshi_2013}
Munshi, F.; Governato, F.; Brooks, A.M.; Christensen, C.; Shen, S.; Loebman, S.; Moster, B.; Quinn, T.; Wadsley, J.
\newblock REPRODUCING THE STELLAR MASS HALO MASS RELATION IN SIMULATED Lambda CDM GALAXIES: THEORY VERSUS OBSERVATIONAL ESTIMATES.
\newblock {\em The Astrophysical Journal} {\bf 2013}, {\em 766},~56.
\newblock {\url{https://doi.org/10.1088/0004-637X/766/1/56}}.

\bibitem[{Conroy}(2013)]{2013ARA&A..51..393C}
{Conroy}, C.
\newblock {Modeling the Panchromatic Spectral Energy Distributions of Galaxies}.
\newblock {\em Annual Review of Astronomy and Astrophysics} {\bf 2013}, {\em 51},~393--455,  \href{http://arxiv.org/abs/1301.7095}{{\normalfont [arXiv:astro-ph.CO/1301.7095]}}.
\newblock {\url{https://doi.org/10.1146/annurev-astro-082812-141017}}.

\bibitem[Kurucz(2011)]{doi:10.1139/p10-104}
Kurucz, R.L.
\newblock Special Issue on the 10th International Colloquium on Atomic Spectra and Oscillator Strengths for Astrophysical and Laboratory Plasmas.
\newblock {\em Canadian Journal of Physics} {\bf 2011}, {\em 89},~417--428,  \href{http://arxiv.org/abs/https://doi.org/10.1139/p10-104}{{\normalfont [https://doi.org/10.1139/p10-104]}}.
\newblock {\url{https://doi.org/10.1139/p10-104}}.

\bibitem[{Levesque} et~al.(2012){Levesque}, {Leitherer}, {Ekstrom}, {Meynet}, and {Schaerer}]{2012ApJ...751...67L}
{Levesque}, E.M.; {Leitherer}, C.; {Ekstrom}, S.; {Meynet}, G.; {Schaerer}, D.
\newblock {The Effects of Stellar Rotation. I. Impact on the Ionizing Spectra and Integrated Properties of Stellar Populations}.
\newblock {\em The Astrophysical Journal} {\bf 2012}, {\em 751},~67,  \href{http://arxiv.org/abs/1203.5109}{{\normalfont [arXiv:astro-ph.SR/1203.5109]}}.
\newblock {\url{https://doi.org/10.1088/0004-637X/751/1/67}}.

\bibitem[{Yan} et~al.(2019){Yan}, {Chen}, {Lazarz}, {Bizyaev}, {Maraston}, {Stringfellow}, {McCarthy}, {Meneses-Goytia}, {Law}, {Thomas}, {Falcon Barroso}, {S{\'a}nchez-Gallego}, {Schlafly}, {Zheng}, {Argudo-Fern{\'a}ndez}, {Beaton}, {Beers}, {Bershady}, {Blanton}, {Brownstein}, {Bundy}, {Chambers}, {Cherinka}, {De Lee}, {Drory}, {Galbany}, {Holtzman}, {Imig}, {Kaiser}, {Kinemuchi}, {Liu}, {Luo}, {Magnier}, {Majewski}, {Nair}, {Oravetz}, {Oravetz}, {Pan}, {Sobeck}, {Stassun}, {Talbot}, {Tremonti}, {Waters}, {Weijmans}, {Wilhelm}, {Zasowski}, {Zhao}, and {Zhao}]{2019ApJ...883..175Y}
{Yan}, R.; {Chen}, Y.; {Lazarz}, D.; {Bizyaev}, D.; {Maraston}, C.; {Stringfellow}, G.S.; {McCarthy}, K.; {Meneses-Goytia}, S.; {Law}, D.R.; {Thomas}, D.;  et~al.
\newblock {SDSS-IV MaStar: A Large and Comprehensive Empirical Stellar Spectral Library{\textemdash}First Release}.
\newblock {\em The Astrophysical Journal} {\bf 2019}, {\em 883},~175,  \href{http://arxiv.org/abs/1812.02745}{{\normalfont [arXiv:astro-ph.IM/1812.02745]}}.
\newblock {\url{https://doi.org/10.3847/1538-4357/ab3ebc}}.

\bibitem[{Lelli} et~al.(2016){Lelli}, {McGaugh}, and {Schombert}]{2016Lelli}
{Lelli}, F.; {McGaugh}, S.S.; {Schombert}, J.M.
\newblock {SPARC: Mass Models for 175 Disk Galaxies with Spitzer Photometry and Accurate Rotation Curves}.
\newblock {\em AJ} {\bf 2016}, {\em 152},~157,  \href{http://arxiv.org/abs/1606.09251}{{\normalfont [1606.09251]}}.
\newblock {\url{https://doi.org/10.3847/0004-6256/152/6/157}}.

\bibitem[Ciotti(2022)]{Ciotti_2022}
Ciotti, L.
\newblock On the Rotation Curve of Disk Galaxies in General Relativity.
\newblock {\em The Astrophysical Journal} {\bf 2022}, {\em 936},~180.
\newblock {\url{https://doi.org/10.3847/1538-4357/ac82b3}}.

\bibitem[Hofmeister and Criss(2017)]{doi:10.1139/cjp-2016-0625}
Hofmeister, A.M.; Criss, R.E.
\newblock The physics of galactic spin.
\newblock {\em Canadian Journal of Physics} {\bf 2017}, {\em 95},~156--166,  \href{http://arxiv.org/abs/https://doi.org/10.1139/cjp-2016-0625}{{\normalfont [https://doi.org/10.1139/cjp-2016-0625]}}.
\newblock {\url{https://doi.org/10.1139/cjp-2016-0625}}.

\bibitem[Sipols and Pavlovich(2020)]{galaxies8020036}
Sipols, A.; Pavlovich, A.
\newblock Dark Matter Dogma: A Study of 214 Galaxies.
\newblock {\em Galaxies} {\bf 2020}, {\em 8}.
\newblock {\url{https://doi.org/10.3390/galaxies8020036}}.

\bibitem[Feng(2020)]{galaxies8010009}
Feng, J.Q.
\newblock Rotating Disk Galaxies without Dark Matter Based on Scientific Reasoning.
\newblock {\em Galaxies} {\bf 2020}, {\em 8}.
\newblock {\url{https://doi.org/10.3390/galaxies8010009}}.

\bibitem[{Milgrom}(1983)]{Milgrom}
{Milgrom}, M.
\newblock {A modification of the Newtonian dynamics - Implications for galaxies}.
\newblock {\em The Astrophysical Journal} {\bf 1983}, {\em 270},~371--389.
\newblock {\url{https://doi.org/10.1086/161131}}.

\bibitem[Wald(1984)]{Wald}
Wald, R.
\newblock {\em General Relativity}; University of Chicago Press, Chicago, IL, USA,  1984.

\bibitem[Tecchiolli(2019)]{universe5100206}
Tecchiolli, M.
\newblock On the Mathematics of Coframe Formalism and Einstein and Cartan Theory. A Brief Review.
\newblock {\em Universe} {\bf 2019}, {\em 5}.
\newblock {\url{https://doi.org/10.3390/universe5100206}}.

\bibitem[Jackson(1999)]{Jack}
Jackson, J.
\newblock {\em Classical Electrodynamics}, 3rd ed.; John Wiley \& Sons, Inc., New Jersey, USA,  1999.

\bibitem[Rindler(2013)]{rindler2013essential}
Rindler, W.
\newblock {\em Essential Relativity: Special, General, and Cosmological}; Springer New York,  2013.

\bibitem[{Casertano}(1983)]{1983MNRAS.203..735C}
{Casertano}, S.
\newblock {Rotation curve of the edge-on spiral galaxy NGC 5907 : disc and halo masses.}
\newblock {\em Monthly Notices of the Royal Astronomical Society} {\bf 1983}, {\em 203},~735--747.
\newblock {\url{https://doi.org/10.1093/mnras/203.3.735}}.

\bibitem[Pomar\`ede et~al.(2020)Pomar\`ede, Tully, Graziani, Courtois, Hoffman, and Lezmy]{Pomarede:2020pme}
Pomar\`ede, D.; Tully, R.B.; Graziani, R.; Courtois, H.M.; Hoffman, Y.; Lezmy, J.
\newblock {Cosmicflows-3: The South Pole Wall}.
\newblock {\em The Astrophysical Journal} {\bf 2020}, {\em 897},~133,  \href{http://arxiv.org/abs/2007.04414}{{\normalfont [arXiv:astro-ph.CO/2007.04414]}}.
\newblock {\url{https://doi.org/10.3847/1538-4357/ab9952}}.

\bibitem[{Schwarzschild}(1954)]{1954AJ.....59..273S}
{Schwarzschild}, M.
\newblock {Mass distribution and mass-luminosity ratio in galaxies}.
\newblock {\em The Astronomical Journal} {\bf 1954}, {\em 59},~273.
\newblock {\url{https://doi.org/10.1086/107013}}.

\bibitem[{Freeman}(1970)]{Freeman}
{Freeman}, K.C.
\newblock {On the Disks of Spiral and S0 Galaxies}.
\newblock {\em The Astrophysical Journal} {\bf 1970}, {\em 160},~811.
\newblock {\url{https://doi.org/10.1086/150474}}.

\bibitem[Binney and Tremaine(2008)]{Binney}
Binney, J.; Tremaine, S.
\newblock {\em Galactic Dynamics}, 2nd ed.; Princeton University Press: Princeton, NJ,  2008.

\bibitem[{Bratek} et~al.(2008){Bratek}, {Ja{\l}ocha}, and {Kutschera}]{2008MNRAS.391.1373B}
{Bratek}, {\L}.; {Ja{\l}ocha}, J.; {Kutschera}, M.
\newblock On the axisymmetric thin disc model of flattened galaxies.
\newblock {\em Monthly Notices of the Royal Astronomical Society} {\bf 2008}, {\em 391},~1373--1383.
\newblock {\url{https://doi.org/10.1111/j.1365-2966.2008.13978.x}}.

\bibitem[Bekenstein(2004)]{PhysRevDBekenstein2004}
Bekenstein, J.D.
\newblock Relativistic gravitation theory for the modified Newtonian dynamics paradigm.
\newblock {\em Phys. Rev. D} {\bf 2004}, {\em 70},~083509.
\newblock {\url{https://doi.org/10.1103/PhysRevD.70.083509}}.

\bibitem[McGaugh(2008)]{McGaugh_2008}
McGaugh, S.S.
\newblock Milky Way Mass Models and MOND.
\newblock {\em The Astrophysical Journal} {\bf 2008}, {\em 683},~137.
\newblock {\url{https://doi.org/10.1086/589148}}.

\bibitem[{Mistele} et~al.(2022){Mistele}, {McGaugh}, and {Hossenfelder}]{2022A&A...664A..40M}
{Mistele}, T.; {McGaugh}, S.; {Hossenfelder}, S.
\newblock {Galactic mass-to-light ratios with superfluid dark matter}.
\newblock {\em Astronomy and Astrophysics} {\bf 2022}, {\em 664},~A40,  \href{http://arxiv.org/abs/2201.07282}{{\normalfont [arXiv:astro-ph.GA/2201.07282]}}.
\newblock {\url{https://doi.org/10.1051/0004-6361/202243216}}.

\bibitem[Loebman et~al.(2014)Loebman, Ivezic', Quinn, Bovy, Christensen, Jurić, Roškar, Brooks, and Governato]{Loebman_2014}
Loebman, S.R.; Ivezic', Z.; Quinn, T.R.; Bovy, J.; Christensen, C.R.; Jurić, M.; Roškar, R.; Brooks, A.M.; Governato, F.
\newblock THE MILKY WAY TOMOGRAPHY WITH SLOAN DIGITAL SKY SURVEY. V. MAPPING THE DARK MATTER HALO.
\newblock {\em The Astrophysical Journal} {\bf 2014}, {\em 794},~151.
\newblock {\url{https://doi.org/10.1088/0004-637X/794/2/151}}.

\bibitem[{Hur{\'e}} and {Hersant}(2011)]{2011A&A...531A..36H}
{Hur{\'e}}, J.M.; {Hersant}, F.
\newblock {The Newtonian potential of thin disks}.
\newblock {\em Astronomy and Astrophysics} {\bf 2011}, {\em 531},~A36,  \href{http://arxiv.org/abs/1104.5079}{{\normalfont [arXiv:astro-ph.IM/1104.5079]}}.
\newblock {\url{https://doi.org/10.1051/0004-6361/201015854}}.

\bibitem[Chatterjee(1987)]{Chatterjee}
Chatterjee, T.
\newblock Potential and force due to thin disk and spherical galaxies.
\newblock {\em Astrophysics and Space Science} {\bf 1987}, {\em 139},~243.

\bibitem[{de Vaucouleurs}(1959)]{1959HDP....53..275D}
{de Vaucouleurs}, G.
\newblock {Classification and Morphology of External Galaxies.}
\newblock {\em Handbuch der Physik} {\bf 1959}, {\em 53},~275.
\newblock {\url{https://doi.org/10.1007/978-3-642-45932-0_7}}.

\bibitem[McGaugh and De~Blok(1998)]{mcgaugh1998testing}
McGaugh, S.S.; De~Blok, W.
\newblock Testing the hypothesis of modified dynamics with low surface brightness galaxies and other evidence.
\newblock {\em The Astrophysical Journal} {\bf 1998}, {\em 499},~66.

\bibitem[McGaugh et~al.(2016)McGaugh, Lelli, and Schombert]{McGaugh2016RAR}
McGaugh, S.S.; Lelli, F.; Schombert, J.M.
\newblock Radial Acceleration Relation in Rotationally Supported Galaxies.
\newblock {\em Phys. Rev. Lett.} {\bf 2016}, {\em 117},~201101.
\newblock {\url{https://doi.org/10.1103/PhysRevLett.117.201101}}.

\bibitem[Navarro(1998)]{JNav}
Navarro, J.
\newblock {The Cosmological Significance of Disk Galaxy Rotation Curves}.
\newblock {\em arxiv:astro-ph/9807084} {\bf 1998}.

\bibitem[Bottema et~al.(2002)Bottema, Pestana, Rothberg, and Sanders]{Bot}
Bottema, R.; Pestana, J.; Rothberg, B.; Sanders, R.
\newblock MOND rotation curves for spiral galaxies with Cepheid based distances.
\newblock {\em Astronomy and Astrophysics} {\bf 2002}, {\em 393},~453.

\bibitem[de~Blok et~al.(2008)de~Blok, Walter, , and Brinks]{Blok1}
de~Blok, W.; Walter, F.; .; Brinks, E.
\newblock High-Resolution Rotation Curves and Galaxy Mass Models from THINGS.
\newblock {\em The Astronomical Journal} {\bf 2008}, {\em 136},~2648.

\bibitem[Gentile et~al.(2011)Gentile, Farnaey, and de~Blok]{Gent}
Gentile, G.; Farnaey, B.; de~Blok, W.
\newblock THINGS about MOND.
\newblock {\em Astronomy and Astrophysics} {\bf 2011}, {\em 527},~A76.

\bibitem[Xue et~al.(2008)]{Xue}
Xue, X.;  et~al.
\newblock {The Milky Way's Circular Velocity Curve to 60 kpc and an Estimate of the Dark Matter Halo Mass from Kinematics of ~2400 SDSS Blue Horizontal Branch Stars}.
\newblock {\em The Astrophysical Journal} {\bf 2008}, {\em 684},~1143--1158,  \href{http://arxiv.org/abs/0801.1232}{{\normalfont [arXiv:astro-ph/0801.1232]}}.
\newblock {\url{https://doi.org/10.1086/589500}}.

\bibitem[{Sofue}(2013)]{Sofue}
{Sofue}, Y., {Mass Distribution and Rotation Curve in the Galaxy}.
\newblock In {\em Planets, Stars and Stellar Systems.~Volume 5: Galactic Structure and Stellar Populations}; Oswalt, T.~D. and Gilmore, G.,  2013; p. 985.
\newblock {\url{https://doi.org/10.1007/978-94-007-5612-019}}.

\bibitem[{Jiao, Yongjun} et~al.(2023){Jiao, Yongjun}, {Hammer, Fran\c{c}ois}, {Wang, Haifeng}, {Wang, Jianling}, {Amram, Philippe}, {Chemin, Laurent}, and {Yang, Yanbin}]{jiao2023detection}
{Jiao, Yongjun}.; {Hammer, Fran\c{c}ois}.; {Wang, Haifeng}.; {Wang, Jianling}.; {Amram, Philippe}.; {Chemin, Laurent}.; {Yang, Yanbin}.
\newblock Detection of the Keplerian decline in the Milky Way rotation curve.
\newblock {\em Astronomy and Astrophysics} {\bf 2023}, {\em 678},~A208.
\newblock {\url{https://doi.org/10.1051/0004-6361/202347513}}.

\bibitem[McGaugh(2019)]{McGaugh_2019}
McGaugh, S.S.
\newblock The Imprint of Spiral Arms on the Galactic Rotation Curve.
\newblock {\em The Astrophysical Journal} {\bf 2019}, {\em 885},~87.
\newblock {\url{https://doi.org/10.3847/1538-4357/ab479b}}.

\bibitem[Lelli et~al.(2016)Lelli, McGaugh, Schombert, and Pawlowski]{Lelli_2016surface}
Lelli, F.; McGaugh, S.S.; Schombert, J.M.; Pawlowski, M.S.
\newblock THE RELATION BETWEEN STELLAR AND DYNAMICAL SURFACE DENSITIES IN THE CENTRAL REGIONS OF DISK GALAXIES.
\newblock {\em The Astrophysical Journal} {\bf 2016}, {\em 827},~L19.
\newblock {\url{https://doi.org/10.3847/2041-8205/827/1/l19}}.

\bibitem[McQuinn et~al.(2019)McQuinn, Boyer, Skillman, and Dolphin]{McQuinn_2019}
McQuinn, K.B.W.; Boyer, M.; Skillman, E.D.; Dolphin, A.E.
\newblock Using the Tip of the Red Giant Branch As a Distance Indicator in the Near Infrared.
\newblock {\em The Astrophysical Journal} {\bf 2019}, {\em 880},~63.
\newblock {\url{https://doi.org/10.3847/1538-4357/ab2627}}.

\bibitem[Kovács(2003)]{10.1046/j.1365-8711.2003.06786.x}
Kovács, G.
\newblock Consistent distances from Baade–Wesselink analyses of Cepheids and RR Lyraes.
\newblock {\em Monthly Notices of the Royal Astronomical Society} {\bf 2003}, {\em 342},~L58--L62,  \href{http://arxiv.org/abs/https://academic.oup.com/mnras/article-pdf/342/4/L58/2830990/342-4-L58.pdf}{{\normalfont [https://academic.oup.com/mnras/article-pdf/342/4/L58/2830990/342-4-L58.pdf]}}.
\newblock {\url{https://doi.org/10.1046/j.1365-8711.2003.06786.x}}.

\bibitem[Oh et~al.(2008)Oh, De~Blok, Walter, Brinks, and Kennicutt]{oh2008high}
Oh, S.H.; De~Blok, W.; Walter, F.; Brinks, E.; Kennicutt, R.C.
\newblock High-resolution dark matter density profiles of things dwarf galaxies: correcting for noncircular motions.
\newblock {\em The Astronomical Journal} {\bf 2008}, {\em 136},~2761.

\bibitem[Željko Ivezić et~al.(2019)Željko Ivezić, Kahn, Tyson, Abel, Acosta, Allsman, Alonso, AlSayyad, Anderson, Andrew, Angel, Angeli, Ansari, Antilogus, Araujo, Armstrong, Arndt, Astier, Éric Aubourg, Auza, Axelrod, Bard, Barr, Barrau, Bartlett, Bauer, Bauman, Baumont, Bechtol, Bechtol, Becker, Becla, Beldica, Bellavia, Bianco, Biswas, Blanc, Blazek, Blandford, Bloom, Bogart, Bond, Booth, Borgland, Borne, Bosch, Boutigny, Brackett, Bradshaw, Brandt, Brown, Bullock, Burchat, Burke, Cagnoli, Calabrese, Callahan, Callen, Carlin, Carlson, Chandrasekharan, Charles-Emerson, Chesley, Cheu, Chiang, Chiang, Chirino, Chow, Ciardi, Claver, Cohen-Tanugi, Cockrum, Coles, Connolly, Cook, Cooray, Covey, Cribbs, Cui, Cutri, Daly, Daniel, Daruich, Daubard, Daues, Dawson, Delgado, Dellapenna, de~Peyster, de~Val-Borro, Digel, Doherty, Dubois, Dubois-Felsmann, Durech, Economou, Eifler, Eracleous, Emmons, Neto, Ferguson, Figueroa, Fisher-Levine, Focke, Foss, Frank, Freemon, Gangler, Gawiser, Geary, Gee, Geha, Gessner,
  Gibson, Gilmore, Glanzman, Glick, Goldina, Goldstein, Goodenow, Graham, Gressler, Gris, Guy, Guyonnet, Haller, Harris, Hascall, Haupt, Hernandez, Herrmann, Hileman, Hoblitt, Hodgson, Hogan, Howard, Huang, Huffer, Ingraham, Innes, Jacoby, Jain, Jammes, Jee, Jenness, Jernigan, Jevremović, Johns, Johnson, Johnson, Jones, Juramy-Gilles, Jurić, Kalirai, Kallivayalil, Kalmbach, Kantor, Karst, Kasliwal, Kelly, Kessler, Kinnison, Kirkby, Knox, Kotov, Krabbendam, Krughoff, Kubánek, Kuczewski, Kulkarni, Ku, Kurita, Lage, Lambert, Lange, Langton, Guillou, Levine, Liang, Lim, Lintott, Long, Lopez, Lotz, Lupton, Lust, MacArthur, Mahabal, Mandelbaum, Markiewicz, Marsh, Marshall, Marshall, May, McKercher, McQueen, Meyers, Migliore, Miller, Mills, Miraval, Moeyens, Moolekamp, Monet, Moniez, Monkewitz, Montgomery, Morrison, Mueller, Muller, Arancibia, Neill, Newbry, Nief, Nomerotski, Nordby, O’Connor, Oliver, Olivier, Olsen, O’Mullane, Ortiz, Osier, Owen, Pain, Palecek, Parejko, Parsons, Pease, Peterson, Peterson,
  Petravick, Petrick, Petry, Pierfederici, Pietrowicz, Pike, Pinto, Plante, Plate, Plutchak, Price, Prouza, Radeka, Rajagopal, Rasmussen, Regnault, Reil, Reiss, Reuter, Ridgway, Riot, Ritz, Robinson, Roby, Roodman, Rosing, Roucelle, Rumore, Russo, Saha, Sassolas, Schalk, Schellart, Schindler, Schmidt, Schneider, Schneider, Schoening, Schumacher, Schwamb, Sebag, Selvy, Sembroski, Seppala, Serio, Serrano, Shaw, Shipsey, Sick, Silvestri, Slater, Smith, Smith, Sobhani, Soldahl, Storrie-Lombardi, Stover, Strauss, Street, Stubbs, Sullivan, Sweeney, Swinbank, Szalay, Takacs, Tether, Thaler, Thayer, Thomas, Thornton, Thukral, Tice, Trilling, Turri, Berg, Berk, Vetter, Virieux, Vucina, Wahl, Walkowicz, Walsh, Walter, Wang, Wang, Warner, Wiecha, Willman, Winters, Wittman, Wolff, Wood-Vasey, Wu, Xin, Yoachim, and Zhan]{Ivezić_2019}
Željko Ivezić.; Kahn, S.M.; Tyson, J.A.; Abel, B.; Acosta, E.; Allsman, R.; Alonso, D.; AlSayyad, Y.; Anderson, S.F.; Andrew, J.;  et~al.
\newblock LSST: From Science Drivers to Reference Design and Anticipated Data Products.
\newblock {\em The Astrophysical Journal} {\bf 2019}, {\em 873},~111.
\newblock {\url{https://doi.org/10.3847/1538-4357/ab042c}}.

\bibitem[McMillan and Binney(2013)]{10.1093/mnras/stt814}
McMillan, P.J.; Binney, J.J.
\newblock {Analysing surveys of our Galaxy ‚Äì II. Determining the potential}.
\newblock {\em Monthly Notices of the Royal Astronomical Society} {\bf 2013}, {\em 433},~1411--1424,  \href{http://arxiv.org/abs/https://academic.oup.com/mnras/article-pdf/433/2/1411/4920807/stt814.pdf}{{\normalfont [https://academic.oup.com/mnras/article-pdf/433/2/1411/4920807/stt814.pdf]}}.
\newblock {\url{https://doi.org/10.1093/mnras/stt814}}.

\bibitem[{Fich} and {Tremaine}(1991)]{1991ARA&A..29..409F}
{Fich}, M.; {Tremaine}, S.
\newblock The mass of the Galaxy.
\newblock {\em Annual Review of Astronomy and Astrophysics} {\bf 1991}, {\em 29},~409--445.
\newblock {\url{https://doi.org/10.1146/annurev.aa.29.090191.002205}}.

\bibitem[Koop et~al.(2024)Koop, Antoja, Helmi, Callingham, and Laporte]{Koop_2024}
Koop, O.; Antoja, T.; Helmi, A.; Callingham, T.M.; Laporte, C.F.P.
\newblock Assessing the robustness of the Galactic rotation curve inferred from the Jeans equations using Gaia DR3 and cosmological simulations.
\newblock {\em Astronomy and Astrophysics} {\bf 2024}, {\em 692},~A50.
\newblock {\url{https://doi.org/10.1051/0004-6361/202450911}}.

\bibitem[Sofue et~al.(2009)Sofue, Honma, and Omodaka]{sofue2009unified}
Sofue, Y.; Honma, M.; Omodaka, T.
\newblock Unified Rotation Curve of the Galaxy—Decomposition into de Vaucouleurs Bulge, Disk, Dark Halo, and the 9-kpc Rotation Dip—.
\newblock {\em Publications of the Astronomical Society of Japan} {\bf 2009}, {\em 61},~227--236.

\bibitem[{Lin} and {Li}(2019)]{2019MNRAS.487.5679L}
{Lin}, H.N.; {Li}, X.
\newblock {The dark matter profiles in the Milky Way}.
\newblock {\em MNRAS} {\bf 2019}, {\em 487},~5679--5684,  \href{http://arxiv.org/abs/1906.08419}{{\normalfont [arXiv:astro-ph.GA/1906.08419]}}.
\newblock {\url{https://doi.org/10.1093/mnras/stz1698}}.

\bibitem[Crézé et~al.(1997)Crézé, Chereul, Bienaymé, and Pichon]{creze1997}
Crézé, M.; Chereul, E.; Bienaymé, O.; Pichon, C.
\newblock The distribution of nearby stars in phase space mapped by Hipparcos: I. The potential well and local dynamical mass,  1997,  \href{http://arxiv.org/abs/astro-ph/9709022}{{\normalfont [arXiv:astro-ph/astro-ph/9709022]}}.

\bibitem[Brownstein and Moffat(2006)]{Brownstein_2006}
Brownstein, J.R.; Moffat, J.W.
\newblock Galaxy Rotation Curves without Nonbaryonic Dark Matter.
\newblock {\em The Astrophysical Journal} {\bf 2006}, {\em 636},~721.
\newblock {\url{https://doi.org/10.1086/498208}}.

\bibitem[{Sanders}(2010)]{2010dmp..book.....S}
{Sanders}, R.H.
\newblock {\em The Dark Matter Problem: A Historical Perspective}; Cambridge University Press,  2010.

\bibitem[Tully et~al.(2014)Tully, Courtois, Hoffman, and Pomar\`ede]{Tully:2014gfa}
Tully, R.B.; Courtois, H.; Hoffman, Y.; Pomar\`ede, D.
\newblock {The Laniakea supercluster of galaxies}.
\newblock {\em Nature} {\bf 2014}, {\em 513},~71,  \href{http://arxiv.org/abs/1409.0880}{{\normalfont [arXiv:astro-ph.CO/1409.0880]}}.
\newblock {\url{https://doi.org/10.1038/nature13674}}.

\bibitem[Naidu et~al.(2022)Naidu, Oesch, van Dokkum, Nelson, Suess, Brammer, Whitaker, Illingworth, Bouwens, Tacchella, Matthee, Allen, Bezanson, Conroy, Labbe, Leja, Leonova, Magee, Price, Setton, Strait, Stefanon, Toft, Weaver, and Weibel]{Naidu_2022}
Naidu, R.P.; Oesch, P.A.; van Dokkum, P.; Nelson, E.J.; Suess, K.A.; Brammer, G.; Whitaker, K.E.; Illingworth, G.; Bouwens, R.; Tacchella, S.;  et~al.
\newblock Two Remarkably Luminous Galaxy Candidates at z [10–12] Revealed by JWST.
\newblock {\em The Astrophysical Journal} {\bf 2022}, {\em 940},~L14.
\newblock {\url{https://doi.org/10.3847/2041-8213/ac9b22}}.

\bibitem[{Misner} et~al.(1973){Misner}, {Thorne}, and {Wheeler}]{MTW}
{Misner}, C.W.; {Thorne}, K.S.; {Wheeler}, J.A.
\newblock {\em Gravitation}; Princeton University Press,  1973.

\bibitem[Schombert et~al.(2018)Schombert, McGaugh, and Lelli]{10.1093/mnras/sty3223}
Schombert, J.; McGaugh, S.; Lelli, F.
\newblock The mass-to-light ratios and the star formation histories of disc galaxies.
\newblock {\em Monthly Notices of the Royal Astronomical Society} {\bf 2018}, {\em 483},~1496--1512,  \href{http://arxiv.org/abs/https://academic.oup.com/mnras/article-pdf/483/2/1496/27089556/sty3223.pdf}{{\normalfont [https://academic.oup.com/mnras/article-pdf/483/2/1496/27089556/sty3223.pdf]}}.
\newblock {\url{https://doi.org/10.1093/mnras/sty3223}}.

\bibitem[Bell and de~Jong(2001)]{BelldYong}
Bell, E.F.; de~Jong, R.S.
\newblock {Stellar Mass-to-Light Ratios and the Tully-Fisher Relation}.
\newblock {\em The Astrophysical Journal} {\bf 2001}, {\em 550},~212--229,  \href{http://arxiv.org/abs/astro-ph/0011493}{{\normalfont [astro-ph/0011493]}}.
\newblock {\url{https://doi.org/10.1086/319728}}.

\bibitem[Lelli et~al.(2017)Lelli, McGaugh, Schombert, and Pawlowski]{Lelli_2017}
Lelli, F.; McGaugh, S.S.; Schombert, J.M.; Pawlowski, M.S.
\newblock One Law to Rule Them All: The Radial Acceleration Relation of Galaxies.
\newblock {\em The Astrophysical Journal} {\bf 2017}, {\em 836},~152.
\newblock {\url{https://doi.org/10.3847/1538-4357/836/2/152}}.

\bibitem[Navarro et~al.(1997)Navarro, Frenk, and White]{NFW}
Navarro, J.; Frenk, C.; White, S.
\newblock The Structure of Cold Dark Matter Halos.
\newblock {\em The Astrophysical Journal} {\bf 1997}, {\em 462},~563.

\bibitem[Wiegart(2010)]{TWiegert}
Wiegart, T.
\newblock Spiral galaxy HI models, rotation curves and kinematic classifications.
\newblock PhD thesis, The University of Manitoba,  2010.

\bibitem[Battaglia et~al.(2006)Battaglia, Fraternali, Oosterloo, , and Sancisi]{Batt}
Battaglia, G.; Fraternali, F.; Oosterloo, T.; .; Sancisi, R.
\newblock HI study of the warped spiral galaxy NGC 5055: a disk/dark matter halo offset?
\newblock {\em Astronomy and Astrophysics} {\bf 2006}, {\em 447},~49.

\bibitem[Fraternali et~al.(2011)Fraternali, Sancisi, , and Kamphuis]{Frat1}
Fraternali, F.; Sancisi, R.; .; Kamphuis, P.
\newblock A tale of two galaxies, light and mass NGC 891 and NGC 7814.
\newblock {\em Astronomy and Astrophysics} {\bf 2011}, {\em http://arxiv.org/abs/1105.3867}.

\bibitem[van~den Bosch and Swaters(2001)]{10.1046/j.1365-8711.2001.04456.x}
van~den Bosch, F.C.; Swaters, R.A.
\newblock Dwarf galaxy rotation curves and the core problem of dark matter haloes.
\newblock {\em Monthly Notices of the Royal Astronomical Society} {\bf 2001}, {\em 325},~1017--1038,  \href{http://arxiv.org/abs/https://academic.oup.com/mnras/article-pdf/325/3/1017/2839453/325-3-1017.pdf}{{\normalfont [https://academic.oup.com/mnras/article-pdf/325/3/1017/2839453/325-3-1017.pdf]}}.
\newblock {\url{https://doi.org/10.1046/j.1365-8711.2001.04456.x}}.

\bibitem[{Navarro} et~al.(2017){Navarro}, {Ben{\'\i}tez-Llambay}, {Fattahi}, {Frenk}, {Ludlow}, {Oman}, {Schaller}, and {Theuns}]{2017MNRAS.471.1841N}
{Navarro}, J.F.; {Ben{\'\i}tez-Llambay}, A.; {Fattahi}, A.; {Frenk}, C.S.; {Ludlow}, A.D.; {Oman}, K.A.; {Schaller}, M.; {Theuns}, T.
\newblock {The origin of the mass discrepancy-acceleration relation in {\ensuremath{\Lambda}}CDM}.
\newblock {\em Monthly Notices of the Royal Astronomical Society} {\bf 2017}, {\em 471},~1841--1848,  \href{http://arxiv.org/abs/1612.06329}{{\normalfont [arXiv:astro-ph.GA/1612.06329]}}.
\newblock {\url{https://doi.org/10.1093/mnras/stx170510.48550/arXiv.1612.06329}}.

\end{thebibliography}
 
%


\PublishersNote{}
\end{adjustwidth}
\end{document}